\documentclass{article}

\usepackage[english]{babel}	

\usepackage{authblk}
\usepackage{amsmath,amssymb}
\usepackage{amsthm} 
\usepackage{mymacros}
\usepackage{wasysym}

\usepackage{circuitikz} 
\usepackage[latin1]{inputenc}
\usepackage[T1]{fontenc}
\usepackage[english]{babel}
\usepackage{graphicx,float }
\usepackage{arydshln,booktabs}
\usepackage[left=2.00cm, right=2.00cm, top=2.00cm, bottom=2.00cm]{geometry}
\usepackage{hyperref}
\usepackage{mathtools} 
\usepackage{xcolor} 
\hypersetup{
    colorlinks=true,
    linkcolor=blue,
    citecolor=green!50!black,
    filecolor=magenta,      
    urlcolor=cyan,
    pdftitle={Digital twin of electrical machines},
    pdfpagemode=FullScreen,
    }

\DeclareFontFamily{U}{mathx}{}
\DeclareFontShape{U}{mathx}{m}{n}{<-> mathx10}{}
\DeclareSymbolFont{mathx}{U}{mathx}{m}{n}
\DeclareMathAccent{\widecheck}{0}{mathx}{"71}

\newcommand{\additionalLoss}{P_{\mathrm{Z}}}
\newcommand{\Amatrix}{A}

\newcommand{\boundarytraction}{\boldsymbol{\tau}_{\mathrm{b}}}
\newcommand{\bulkvisc}{\zeta_1}
\newcommand{\chargeDensityLag}{\rho_\mathrm{c}}
\newcommand{\chargePotentialLag}{\boldsymbol{D}}

\newcommand{\currentDensityLag}{\boldsymbol{J}}

\newcommand{\deformationtensor}{F}

\newcommand{\diagMatrix}{\Lambda_0}

\newcommand{\discForceVec}{\boldsymbol{f}}
\newcommand{\dispVec}{\boldsymbol{s}}

\newcommand{\elasticity}{\mathbb{H}}
\newcommand{\elasticityHat}{\widetilde{\elasticity}}
\newcommand{\electricConductivity}{\sigma}
\newcommand{\electricFieldLag}{\boldsymbol{E}}
\newcommand{\electricFluxLinkage}{\psi_{\mathrm{e}}}

\newcommand{\entropy}{s}

\newcommand{\firstLame}{\lambda_1}
\newcommand{\forceField}{\boldsymbol{F}}

\newcommand{\fourthorderidentity}{\mathbb{I}}
\newcommand{\fPiola}{P}
\newcommand{\fPiolaConsti}{\fPiola_{\mathrm{c}}}
\newcommand{\freeenergy}{\psi}
\newcommand{\freeenergyConsti}{\freeenergy_{\mathrm{c}}}

\newcommand{\frictionCoeff}{\mu_{\mathrm{f}}}
\newcommand{\frictionLoss}{P_{\mathrm{R}}}

\newcommand{\gyroscopicMatrix}{G}

\newcommand{\heatFlux}{\boldsymbol{q}}

\newcommand{\heatsource}{r}

\newcommand{\intEnergy}{u}
\newcommand{\ironCoreLoss}{P_{\mathrm{Fe}}}

\newcommand{\JouleHeat}{\heatsource_{\mathrm{J}}}
\newcommand{\KG}{K_\mathrm{G}}
\newcommand{\lag}{\boldsymbol{x}}
\newcommand{\lagComponent}{x}

\newcommand{\magneticFieldLag}{\boldsymbol{B}}
\newcommand{\magneticHField}{\boldsymbol{H}}

\newcommand{\motion}{\boldsymbol{\phi}}

\newcommand{\nPhases}{N_{\mathrm{p}}}
\newcommand{\ODEMassMatrix}{E}

\newcommand{\outwardnormal}{\boldsymbol{\nu}}

\newcommand{\permanentMagneticFluxLinkage}{\psi_{\mathrm{m}}}
\newcommand{\permittivity}{\epsilon}

\newcommand{\powerSourceVec}{\boldsymbol{P}}

\newcommand{\Refconfig}{\mathcal{X}}
\newcommand{\rhoref}{\rho}

\newcommand{\Robinvalue}{\gamma_{\mathrm{b}}}

\newcommand{\rotMat}{H}
\newcommand{\rotorLoss}{P_{\mathrm{w}2}}
\newcommand{\rotVel}{\omega}
\newcommand{\secondLame}{\lambda_2}
\newcommand{\shearvisc}{\zeta_2}
\newcommand{\smalldisp}{\boldsymbol{u}}
\newcommand{\smallforce}{\widecheck{\forceField}}
\newcommand{\specificHeat}{c}
\newcommand{\sPiola}{S}
\newcommand{\state}{\boldsymbol{z}}
\newcommand{\statorLoss}{P_{\mathrm{w}1}}
\newcommand{\surface}{S}
\newcommand{\surfaceElement}{\boldsymbol{\surface}}
\newcommand{\synchronousSpeed}{n_{\mathrm{s}}}
\newcommand{\Te}{T_\mathrm{e}}
\newcommand{\temperature}{\vartheta}

\newcommand{\tempVec}{\state}

\newcommand{\thermalConductivity}{\kappa}

\newcommand{\TL}{T_\mathrm{L}}
\newcommand{\torqueofinertia}{\Theta}

\newcommand{\totalCurrent}{\boldsymbol{I_{\mathrm{T}}}}

\newcommand{\totalLoss}{P_{\mathrm{V}}}
\newcommand{\transposer}{\mathbb{T}}

\newcommand{\vectorPotential}{\boldsymbol{A}}
\newcommand{\velocity}{\boldsymbol{v}}

\newcommand{\viscosity}{\mathbb{V}}
\newcommand{\viscosityHat}{\widetilde{\viscosity}}

\newcommand{\boundaryheatflow}{q_{\mathrm{b}}}
\newcommand{\boundarymotion}{\motion_{\mathrm{b}}}
\newcommand{\boundarytemperature}{\temperature_{\mathrm{b}}}

\newcommand{\inp}[1]{\left\langle #1\right\rangle}

\newcommand{\ipF}[1]{\inp{#1}_\mathrm{F}}

{\left(\begin{smallmatrix}}
{\end{smallmatrix}\right)}

{\left[\begin{smallmatrix}}
{\end{smallmatrix}\right]}

\newtheorem{theorem}{Theorem}
\newtheorem{remark}[theorem]{Remark}

\numberwithin{theorem}{section}

\providecommand{\keywords}[1]
{
  \small	
  \textbf{\textit{Keywords---}} #1
}

\allowdisplaybreaks 

\title{Hierarchical modeling for an industrial implementation of a Digital Twin for electrical drives}

\author[1]{Karim Cherifi}
\author[1]{Philipp Schulze}
\author[1]{Volker Mehrmann}
\affil[1]{Institute of Mathematics, Technische Universit\"at Berlin, Berlin, Germany. \texttt{\{cherifi,pschulze,mehrmann\}@math.tu-berlin.de}}

\author[2]{Leo Go{\ss}lau}
\affil[2]{Innomotics GmbH, Berlin, Germany. \texttt{leo.gosslau@innomotics.com}}
\author[3]{Pascal L\"unnemann}
\affil[3]{Fraunhofer-Institut f\"ur Produktionsanlagen und Konstruktionstechnik IPK, Berlin, Germany. \texttt{pascal.luennemann@ipk.fraunhofer.de}}

\begin{document}
\maketitle

\begin{abstract}
Digital twins have become popular for their ability to monitor and optimize a process or a machine, ideally through its complete life cycle using simulations and sensor data.  In this paper, we focus on the challenge  of accurate and real-time simulations for digital twins in the context of \emph{electrical machines}. 

To build such a digital twin involves not only computational models for the electromagnetic aspects, but also mechanical and thermal effects need to be taken into account.
We address  mathematical tools that can be employed to carry out the required simulations based on physical laws as well as surrogate or data-driven models. 
One of those tools is a model hierarchy of very fine to very coarse models as well as reduced order models  for obtaining real-time simulations.

The required software tools to carry out simulations in the digital twin are also discussed.
The simulation models are implemented in a pipeline that allows for the automatic modeling of new machines and the automatic configuration of new digital twins.
Finally, the overall implemented digital twin is tested and implemented in a physical demonstrator.  
\end{abstract}


\section{Introduction: Numerical simulation and requirements for a digital twin}
\label{sec:intro}

When an  electrical machine is equipped with a large number of sensors, then it produces a lot of data that can be collected, stored, and used \cite{IqbMR21,Mel18}. 
A digital twin is a tool that uses this data combined with numerical simulations to accompany  its physical twin in an optimal way, ideally throughout its life cycle \cite{RasO20,BouVC20,FalK21}. 
By means of the phenomenal advancements in both hardware and software in recent years in terms of data processing, compute power, and data-driven modeling methods, digital twins are becoming a reality that can be useful in many ways. 
This includes the following functions: storage of data history and feedback to design, data based modeling, documentation and design optimization, condition monitoring, and predictive maintenance. 

A digital twin can be used to continuously monitor the behavior of the machine and detect any deviations in the operating conditions of individual components or of the overall system. 
In addition, a digital twin can improve predictive maintenance by continuously predicting failures of system components using detailed life-time simulations of the critical components in the machine. 
For this, the simulation should be adapted to the operating and usage conditions that may affect the original model. 
This can be done using data-driven methods and data assimilation. 
A successful detection and warning in the event of abnormal behavior allows the replacement of faulty components before their breakdown and can prevent damage to other components.
Some of these functions need accurate simulation models that can replicate and predict the operation of the physical twin in real-time. 
The simulation models can also be used to optimize the machine parameters for the respective operating environment based on the knowledge gained from operation.
For example, motor-specific monitoring of the efficiency under parameter variations allows to identify optimized operating points online while the machine is running or offline before turning on the machine based on the system parameters and stored data.

In this context, accurate simulations that can be obtained in real-time are required by some functions of the digital twin. 
However, in most conventional simulation software, it is not possible to obtain a simulation in real-time, thus the models have to simplified. 
For safety reasons, these simplifications and resulting decisions have to be understood, justified, and documented. 
This is one of the main challenges of digital twins as discussed in \cite{RasO20}.
One example of a mathematically sound simplification method that can be used is model order reduction (MOR), see for instance \cite{HarHW18} for a corresponding discussion in the context of digital twins and \cite{BenGQRSM21a,BenGQRSM21b,BenGQRSM21c} for a general overview.
One may, however, also rely on data-driven surrogate models obtained from measurement data \cite{FriFLM22}. 

\begin{figure}[H]
        \centering
        \includegraphics[scale=0.35]{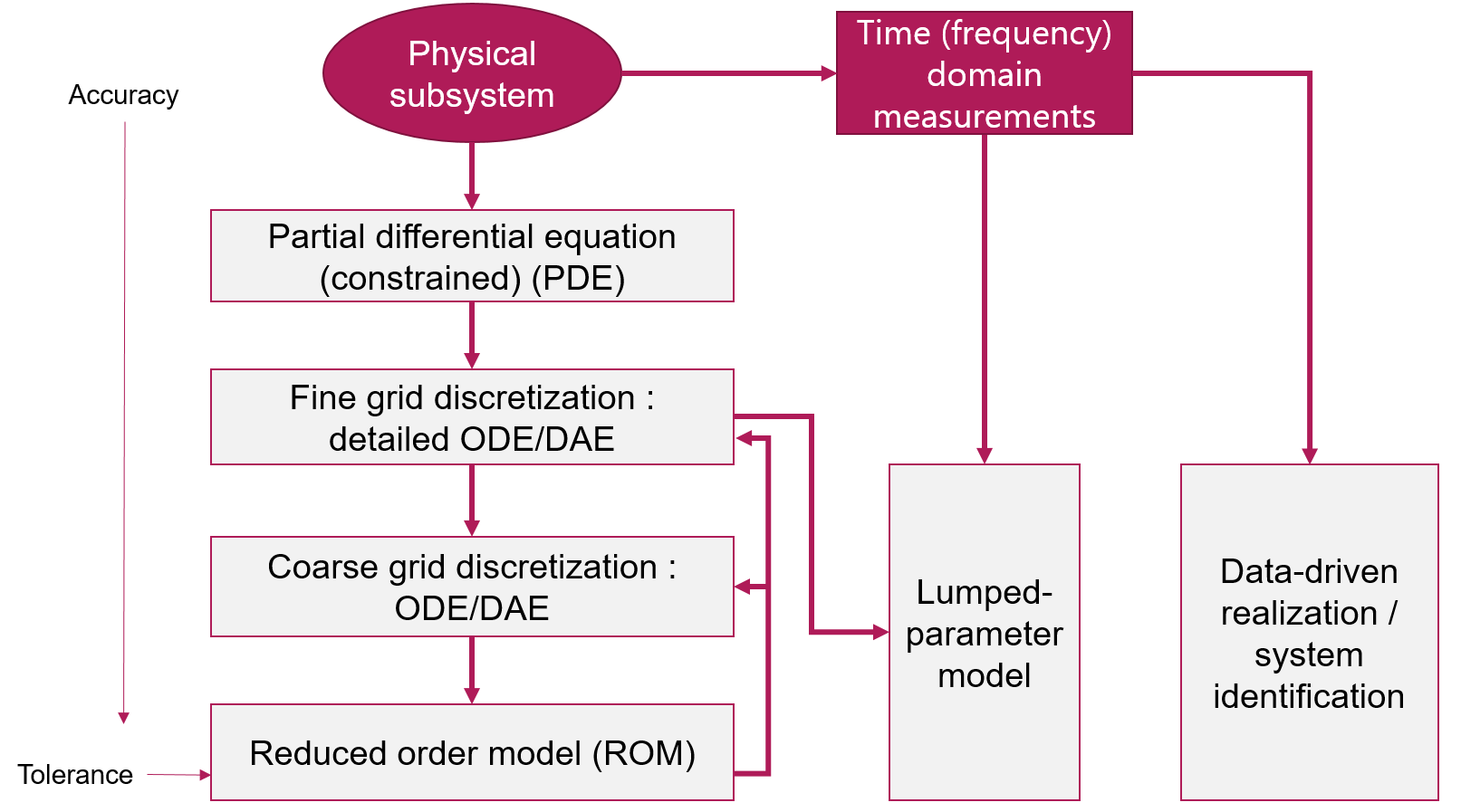}
        \caption{Catalogue of simulation models including a system hierarchy.}
        \label{fig:Hiearchy}
\end{figure}

Altogether, the availability of different models of different scales and accuracy allows to build a catalogue, or hierarchy  of models from detailed complex to simple (and faster) models as illustrated in Figure~\ref{fig:Hiearchy}. 

The physical system is typically described  by a partial differential equation (PDE) together with boundary conditions and may include some algebraic constraints, arising from interconnection and interface conditions or balance laws.

The PDE is then typically discretized in space using a grid hierarchy of coarse to very fine grids that allow for accuracy adjustment. Space discretization results in a system of ordinary differential equations (ODEs) or differential-algebraic equations (DAEs) if we have constraints. 

The resulting dynamical model can be further simplified using reduced order modeling. In this way, the accuracy of the resulting simulation can be adjusted based on a user-defined tolerance used in the model reduction algorithm. Then, depending on the application, one can use a suitable model based on given accuracy and computing-time requirements.

On the other hand, one can describe the dynamics of the physical system by means of a data-driven realization using measurements from the physical system. A data-driven realization can also be used to derive a predefined lumped parameter model. Alternatively, the lumped parameters can be estimated from the discretized PDE.
However, these methods typically do not provide any guarantees about the estimation of the error.

In addition, we also require the simulation to be adaptive to the electrical machine specifications and parameters. For this, we need a simulation setup, where a new simulation can be obtained if some part of the electrical machine  is modified or the electrical machine itself is e.g.~exchanged in a factory. This is particularly important when dealing with high power electric motors that can have power up to 100 MW \cite{RyuE20}. These motors are generally custom-made and thus models have to be adapted to each new motor produced. In addition, one has to consider effects that may affect its life cycle.
For example, the vibrations that result from the operation of these huge machines can end up damaging the machine and shorten its lifetime \cite{thoD99,scheffer2004}. 

The simulation models that we consider in the construction of  a digital twin for an electrical machine can be divided into three main model families: electrical models, mechanical models and thermal models. 
In this paper we derive the basic equations describing the three model families and how these can be incorporated in a digital twin practically, cf.~sections~\ref{sec:electricalSim}--\ref{sec:thermalSimulation}.
In addition, we identify the main challenges that one is faced with in the implementation of a digital twin for electrical machines.
In particular, we discuss the challenge of using simulations within the general software architecture of a digital twin in section \ref{sec:software}. The mathematical models are then used in a pipeline for the automatic configuration of digital twins within the general twin architecture. Finally, in section \ref{sec:demonstrator}, we discuss the implementation of the digital twin in a physical demonstrator.

\paragraph*{Mathematical notation} 

Throughout this manuscript, we use bold font for vectors and vector-valued functions.
We denote the set of real numbers by $\R$, the standard Euclidean space of dimension $n$ by $\R^n$, and the Euclidean inner product of two vectors $\boldsymbol{v},\boldsymbol{w}\in\R^n$ by $\boldsymbol{v}\cdot \boldsymbol{w}$.
Furthermore, we use $\R^{m,n}$ for the space of $m\times n$ matrices and $\R^{n_1,n_2,n_3,n_4}$ for fourth-order tensors of size $n_1\times n_2\times n_3\times n_4$.
The tensor product between two matrices $A\in\R^{m,n}$ and $B\in\R^{p,q}$ is denoted by $A\otimes B \in\R^{m,n,p,q}$ and defined via $[A\otimes B]_{ijkl} \vcentcolon= [A]_{ij}[B]_{k\ell}$ for $i=1,\ldots,m$, $j=1,\ldots,n$, $k=1,\ldots,p$, and $\ell=1,\ldots,q$.
Finally, for the Frobenius inner product of two matrices $A,B\in \R^{m,n}$ we use the notation $\ipF{A,B} \vcentcolon= \trace{A^\top B}$, where $\trace{\cdot}$ denotes the trace.

\section{Electrical (Electromagnetic) simulation}
\label{sec:electricalSim}

In real world digital twins, we ideally require a hierarchy of models, where the appropriate model is chosen based on a trade-off between simulation accuracy and computing-time. 
In the electromagnetic simulation, we start with the most fundamental model based on the Maxwell equations up to the simplest surrogate models based on equivalent circuits of the electrical machine.
Although most of the following equations hold for all rotating machines, there are still some significant differences between synchronous and asynchronous induction machines. In the following, as a use case we focus on induction motors when appropriate.

\subsection{Fundamental equations and finite element simulation}
\label{sec:electromagnetics_model}

In order to study the transient time-domain behavior of rotating machines, we consider the Maxwell equations for homogeneous isotropic materials described by 
\begin{subequations}
\label{eq:MaxwellEq}
\begin{align}
    \nabla \cdot \electricFieldLag  & = \frac{\chargeDensityLag}{{\permittivity }}, \label{eq:MaxwellEq1} \\ 
    \nabla \cdot \magneticFieldLag  & = 0 \label{eq:MaxwellEq2}, \\ 
    \nabla \times \electricFieldLag  & = -\frac{\partial \magneticFieldLag}{\partial t} \label{eq:MaxwellEq3}, \\ 
    \nabla \times \magneticFieldLag  & = \mu(\currentDensityLag+{\permittivity }\frac{\partial \electricFieldLag}{\partial t}), \label{eq:MaxwellEq4}
\end{align}
\end{subequations}
where $\electricFieldLag$ is the electric field, $\magneticFieldLag$ the magnetic flux density, $\currentDensityLag$ the total electric current density, and $\chargeDensityLag$ the total electric charge density.
Furthermore, $\permittivity$ and $\mu$ are the permittivity and the permeability, respectively.
These depend on the material and in free space they just reduce to the vacuum permittivity ${\permittivity }_{0}$ and the vacuum permeability ${\mu}_{0}$, respectively.
We emphasize that \eqref{eq:MaxwellEq} does not take into account any effects resulting from the motion of the material, i.e., the coupling to the mechanics via velocity-dependent effects is neglected in this section.

\begin{figure}[H]
        \centering
        \includegraphics[scale=0.45]{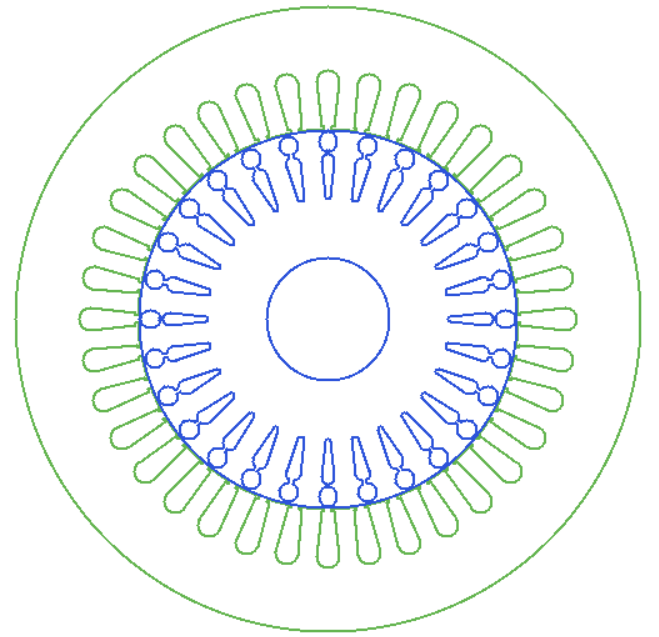}
        \caption{Example 2D configuration of an induction machine in \textsc{Ansys} software}
        \label{fig:Ansys1}
\end{figure}

Although the Maxwell equations can be used in various ways in order to study dynamics of rotating machines, we present below how they are most commonly used and, specifically, how they are employed in commercial software such as \textsc{Ansys Maxwell}. 
One example is the configuration shown in Figure~\ref{fig:Ansys1}.
 
 The following equations focus on the 2D case and are derived mainly from the references \cite{Salon1995,IqbMR21,Bia05}. The 3D case is not discussed but most of the concepts discussed in what follows can be generalized to the 3D case. 
 
First, we focus on the model of the electrical system. The solution for the electric potential $\Phi (\lagComponent_1,\lagComponent_2)$ in a plane represented by  vectors $\lagComponent_1$ and $\lagComponent_2$ (the 2D case) is derived from the Maxwell equation \eqref{eq:MaxwellEq}, resulting in the equation
\begin{equation}
    \label{eq:ElecPot}
    \nabla \cdot({\permittivity}\nabla \Phi )=-\chargeDensityLag.
\end{equation}
This equation is solved using finite element methods (FEM) for a scalar potential function $\Phi$. For more details we refer to \cite{Salon1995}.
\begin{remark}
    In the special case of electrostatics, given the potential function $\Phi$, one can compute the the static field $\electricFieldLag$ as 
    \begin{equation}
    \electricFieldLag =- \nabla\Phi. \label{eq:E}
    \end{equation}
\end{remark}

Given the electric potential and the resulting static field $\electricFieldLag$, one can compute the current density $\currentDensityLag$, the electric flux density $\chargePotentialLag$ and electric flux linkage $\electricFluxLinkage$ in a surface $\surface$ as
\begin{subequations}
\label{eq:ElecQuant}
\begin{align}
    \currentDensityLag & ={\electricConductivity}\electricFieldLag \label{eq:J}, \\
    \chargePotentialLag & ={\permittivity}\electricFieldLag \label{eq:D}, \\ 
    \electricFluxLinkage & =\int_{\surface}^{} {\electricFieldLag\cdot \mathrm{d}\surfaceElement}, \label{eq:psi}
\end{align}
\end{subequations}
where $\electricConductivity$ is the conductivity specific to the material. The total current flowing in the conductor $\totalCurrent$  is then equal to the integral of the current density over the cross-section of the conductor $\Omega$ resulting in the equation
\begin{equation}
    \label{eq:TotalI}
    \totalCurrent = \int\limits_{\Omega }{\currentDensityLag d\Omega }.
\end{equation}

From the electric field, the generated mechanical power and resulting torque are derived, see for instance \cite{CoM84}.

Another quantity of interest is the magnetic vector potential $\vectorPotential$ that is used to compute the magnetic flux density $\magneticFieldLag$, the magnetic field $\magneticHField$ (for linear materials), and the magnetic flux linkage $\psi_i$ in each coil as presented in

\begin{subequations}
\label{eq:MagQuant}
\begin{align}
    \magneticFieldLag & =\nabla \times \vectorPotential  \label{eq:B}, \\ 
    \magneticHField & =\frac{\magneticFieldLag}{\mu} \label{eq:H}, \\ 
    \psi_i & =L_{ii} I_i+\sum\limits_{j=1}^{n}{L_{ij} I_j} +\permanentMagneticFluxLinkage, \label{eq:psi_i}
\end{align}
\end{subequations}
where $L_{ii}$ is the self inductance in a single coil, $L_{ij}$ is the mutual inductance between the coils and $\permanentMagneticFluxLinkage$ is the permanent magnetic flux linkage. 
The derivation of the magnetic flux linkage is discussed in detail in \cite{MonH16}, where the infinite dimensional system is discretized and solved efficiently using model order reduction. 
The reduction of the model size for the model resulting from the discretization of the Maxwell equations is mandatory in the context of a digital twin, to achieve real-time computations and real-time adaptability. 
The work done in \cite{MonH16} represents  an example on how to increase the speed of the simulation.

In the static or quasi-stationary case, the magnetic vector potential $\vectorPotential$ which is the quantity of interest here, can be obtained 
from the Maxwell equations resulting in the diffusion equation
\begin{equation}
    \label{eq:MagneticPot}
    \currentDensityLag={\nabla}\times(\frac{1}{\mu}{\nabla}\times\vectorPotential),
\end{equation}
where $\vectorPotential$ is the magnetic vector potential and $\currentDensityLag$ is the current density field. Here $\vectorPotential$ and $\currentDensityLag$ are assumed to be $\lagComponent_3$-directed and independent of $\lagComponent_3$ as they are a result of the $\lagComponent_1$- and $\lagComponent_2$-directions \cite{Salon1995}. 
Given an excitation $\currentDensityLag$, the FEM solver finds the solution of \eqref{eq:MagneticPot} for $\vectorPotential$. Typically $\currentDensityLag$ is considered to be having three parts: one due to the applied source, another due to the induced electric field produced by time-varying magnetic flux, and the third due to motion-induced voltage. This introduces a time varying term for $\vectorPotential$ and describes the problem of the eddy current effects. Eddy currents are currents induced within conductors by a changing magnetic field in the conductor. 
The eddy current will itself create a magnetic field that causes energy loss in the rotating machine. 
In the case of induction machines, eddy currents are responsible for producing the main torque. Eddy currents and corresponding equations describe the coupling between the electrical and the magnetic field.

In order to compute the eddy currents, we start by rewriting the Maxwell equation \eqref{eq:MaxwellEq3} using equation \eqref{eq:B} as
\begin{equation}
    \label{eq:E11}
  \nabla \times \electricFieldLag  = -\frac{\partial \nabla \times \vectorPotential}{\partial t},\\
\end{equation}
resulting in
\begin{equation}
    \label{eq:E12}
    \electricFieldLag =- \frac{\partial \vectorPotential}{\partial t}-\nabla \Phi. \\
\end{equation}
This reformulation results in an equation that describes the coupling between electric and magnetic fields and their mutual effects based on \eqref{eq:J}, \eqref{eq:MagneticPot}, and \eqref{eq:E12} as
\begin{equation}
    \label{eq:Eddy1}
    \nabla \times (\frac{1}{\mu} \nabla \times  \vectorPotential) =-\electricConductivity \frac{\partial \vectorPotential}{\partial t}-\electricConductivity\nabla \Phi. \\
\end{equation}

\begin{figure}[H]
        \centering
        \includegraphics[scale=0.4]{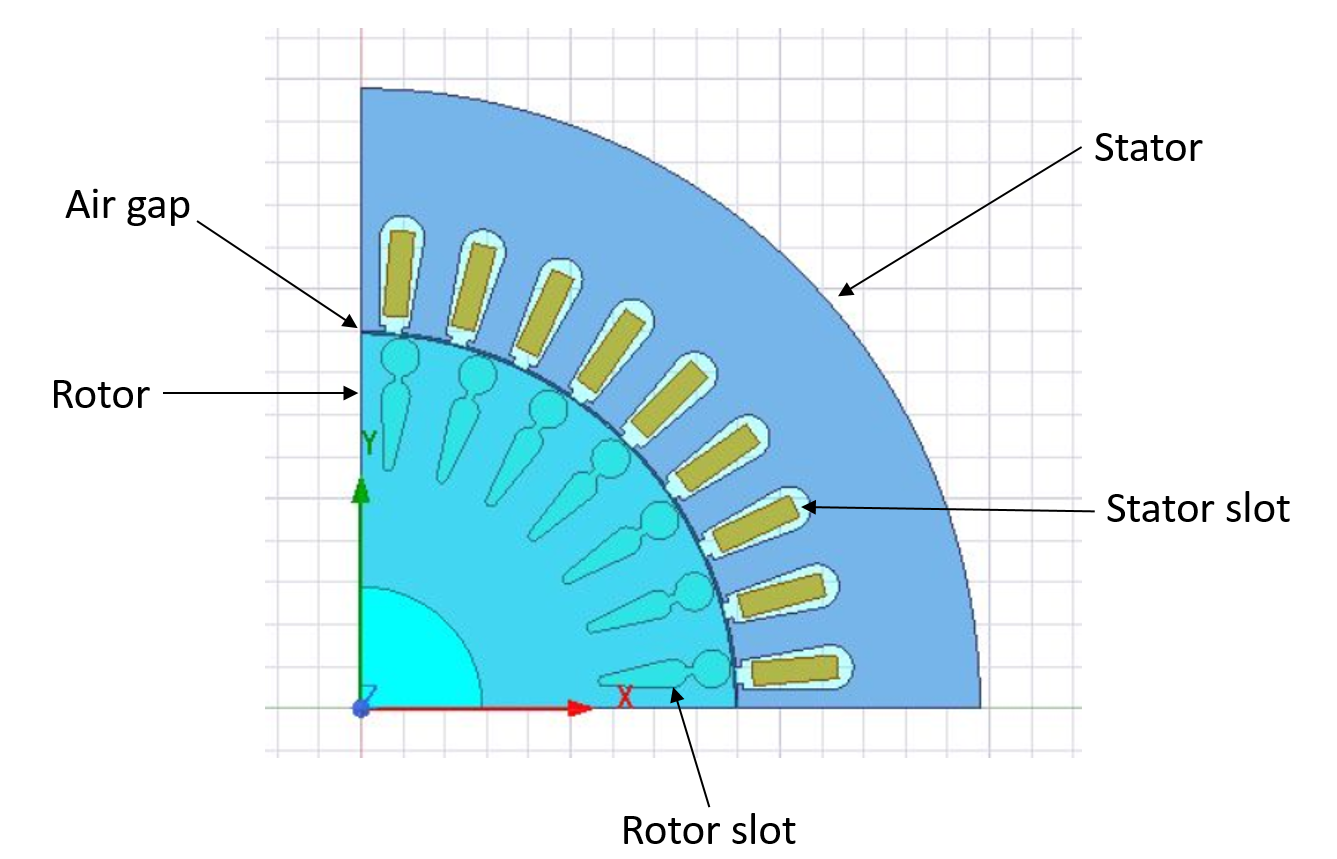}
        \caption{Example of an \textsc{Ansys Maxwell} simulation exploiting symmetry of an induction machine}
        \label{fig:Ansys2}
\end{figure}

This approach is discussed in more detail in \cite{Szu01}. In the 2D case, only the $\lagComponent_1\lagComponent_2$-plane is considered taking into account the symmetry of the machine. 
In the electrical engineering literature, the $\lagComponent_1\lagComponent_2$-plane is commonly referred to as the $xy$-plane while $\lagComponent_3$ is referred to as $z$ as shown in Figure~\ref{fig:Ansys2}. 
As shown in \cite{Bia05}, due to the symmetry assumption, the flux density and magnetic field strength vectors have only non-zero components in $\lagComponent_1$- and $\lagComponent_2$-direction, i.e., $\magneticFieldLag=(B_1,B_2,0)$ and $\magneticHField=(H_1,H_2,0)$. 
The resulting magnetic vector potential and the current density vectors have an $\lagComponent_3$-axis component only, which are $\vectorPotential=(0,0,A_3)$ and $\currentDensityLag=(0,0,J_3)$ respectively.

When applying equations \eqref{eq:MagneticPot} and \eqref{eq:Eddy1} in the context of a rotating electrical machine, one has to consider boundary conditions. This is discussed extensively in \cite{Bia05}. In the special case of asynchronous induction motors, the geometric and magnetic symmetry allows to limit the analysis domain to a portion of the motor section as shown in Figure~\ref{fig:Ansys2}.

This portion is primarily determined by the stator and rotor slot numbers, as well as the stator winding. Another important parameter is the number of magnetic pole pairs considered. These are  periodically arranged, so analysis can be carried out only on one pole pair. 

When a stator symmetry exists in its geometry, for example in the poles of the machine, i.e.~the number of slots is a multiple of the number of poles, the study can be carried out only on one single pole. 
There are usually no problems with geometric symmetry, especially if the slot number is a multiple of the pole number. On the other hand, due to the stator winding, special attention must be paid to magnetic symmetry.
The analysis domain can be condensed by imposing appropriate periodic boundary conditions. 
In addition, special operating conditions allow Dirichlet or Neumann boundary conditions to be assigned \cite{Bia05}. Nevertheless, we consider here the general case in normal operating condition.

A Dirichlet boundary condition $A_3=0$ is assigned to the stator's external circumference, causing the flux lines to remain confined within the stator yoke. However, in case of high saturation, this may be disadvantageous. In that case, it is better to have some air outside the yoke in the model. 
Furthermore, because the shaft is generally excluded from the analysis domain, the Dirichlet boundary condition is also applied between the shaft and the rotor lamination.

Considering the boundary conditions in the context of an asynchronous machine and both geometric and magnetic symmetry conditions, Dirichlet boundary conditions are assigned on the external boundary of the stator and Neumann boundary conditions $\frac{\partial A_3}{\partial \outwardnormal}=0$, in the normal direction $\outwardnormal$ with respect to the $\lagComponent_1 \lagComponent_2$-plane in our context, are assigned on the remaining boundary of the sector constituted of either one pole, a pair of poles or a set of pole pairs. 

\begin{figure}[H]
        \centering
\includegraphics[scale=0.6]{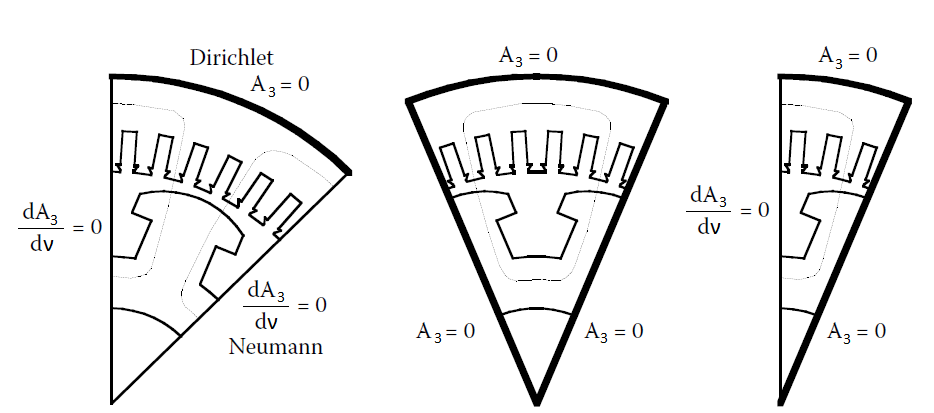}
        \caption{Boundary conditions types considered for the no-load simulation of an electrical machine \cite{Bia05}}
        \label{fig:BC}
\end{figure}

Finally, for other types of electrical machines and also in a magnetically asymmetric context, one can still use periodic boundary conditions assigned along the lateral boundary of the sector. When these are taken into account, the remaining external circumference is subjected to Dirichlet boundary conditions. These types of boundary conditions are illustrated in Figure \ref{fig:BC}.

In terms of initial conditions, the machine is typically simulated starting from rest with an assigned current or current density assigned to specific coils.

More details on space and time discretization are discussed in \cite{Bia05,Salon1995} along with solution methods for the resulting discretized finite dimensional system. 
The solution of equation \eqref{eq:Eddy1} requires a discretization scheme \cite{ManMM08,Isl10}. 
The computation time for a complex problem may take a few hours especially in the 3D case. 
Such computing times cannot be tolerated in a digital twin in real operation. For this, it is necessary to use model reduction to be closer to the real-time requirements. 
Some work in this direction has been proposed \cite{Cod15,SatI13,MueSKNH21} using for example proper orthogonal decomposition (POD) \cite{SatI13} or the Gappy POD and the discrete empirical interpolation method (DEIM) \cite{MueSKNH21}. 

Model reduction was also used in non-symmetric problems for rotating machines \cite{BonLRS18} when small perturbations in the geometry or the material parameters are introduced during the mass production process and have to be taken into account in the model treated by the digital twin for the specific electrical machine used in the factory. 

Model uncertainty is another challenge that has to be taken into account in digital twins. 
This is typically done by incorporating the data gathered from the physical twin to update the parameters in the model of the digital twin. An application in this direction can be found in \cite{KapKHTW20}.

In terms of simulation software, \textsc{Ansys Maxwell} is one of the industrial standards in solving the equations presented in this subsection. 
It makes the design and analysis of these electrical machines easy with preset models of different motor classes and provides also a readily available visualization interface. 
However, the software lacks flexibility and transparency in terms of model reduction that is an essential part of digital twins. More recently, \textsc{Ansys} presented \textsc{Ansys Twin Builder} to answer this question. However, an effort should  still be made to use the wide range of model reduction techniques existing in the literature, especially for coupled systems. In addition, a model hierarchy is also required to have the best model for a specific application, see for instance \cite{BesDDCGT04,DomHLMMT21,LiFW21} for some examples of model hierarchies in different application fields from engineering and physics. In a digital twin where model uncertainty and fast computations have to be taken into account, \textsc{Ansys} has to be more open and flexible concerning model extraction and manipulation. 

Open-source software, such as e.g.~\textsc{FEniCS}, can also be used to solve the equations mentioned above. The mathematical model would then be available and could be adapted, reduced, or coupled with other systems. Although this gives more freedom, it requires the implementation of the models for different types of machines and their coupled models. 
In addition, 
\textsc{Ansys}, still has more attractive visualization tools than most open-source software implementations for electrical machines.

To summarize, in the simulation of electrical rotating machines, the following assumptions are generally used:
\begin{itemize}
    \item Linear materials are assumed to obtain the linear relations in equations \eqref{eq:ElecQuant} and \eqref{eq:MagQuant} that include polarization and magnetization microscopic effects.
    \item When considering only electromagnetic simulations, the mechanical effects such as velocity-dependent effects and deformation are neglected.
    \end{itemize}

\subsection{Lumped parameter model: Equivalent circuit model}
\label{sec:equ_ckt}
It is popular in electrical engineering to use lumped-parameter and surrogate models. 
These parameters are represented by impedance values, voltage sources and current sources that constitute an equivalent circuit of the machine. 
These calculation models are extremely fast but do not have any guarantees on error bounds. This is the reason why these models are placed at the bottom of our hierarchy of models. The surrogate models mainly used in steady state electromagnetic simulation are based on equivalent circuit models of the electrical machine. These models are specific to the type of machines being studied. For asynchronous AC machines, the equivalent circuit can be derived through a series of simplifications and approximations \cite[Ch.~4]{Mel18}.

\begin{figure}
 \centering
      \begin{circuitikz}[european]
		\draw node[ground]{} (0,0) to[sinusoidal voltage source, l=$U_1$](0,3);
		\draw (0,3) to [short,i=$I_1$] (0.8,3) to[R=$R_1$] (2.2,3)--(4,3);
		\draw (4,3) to [R=$X_{1}$] (5.2,3)--(6,3) ;
		\draw (6,3) -- (6,2.8) to[R=$X_3$] (6,0) node[ground]{};
		\draw (6,3) to [short,i=$I'_2$] (6.8,3) to[R=$X'_{2}$] (8,3) ;
		\draw (8,3) -- (9,3) to[R=$\frac{R'_2}{s}$] (10.2,3) -- (11,3) ;
		\draw (11,3) -- (11,0) node[ground]{} ;
		\node (a) at (11.4,3) {$U'_2$};
	\end{circuitikz}
	\caption{Equivalent circuit model of an induction machine}
	\label{fig:EqCKT}
\end{figure}
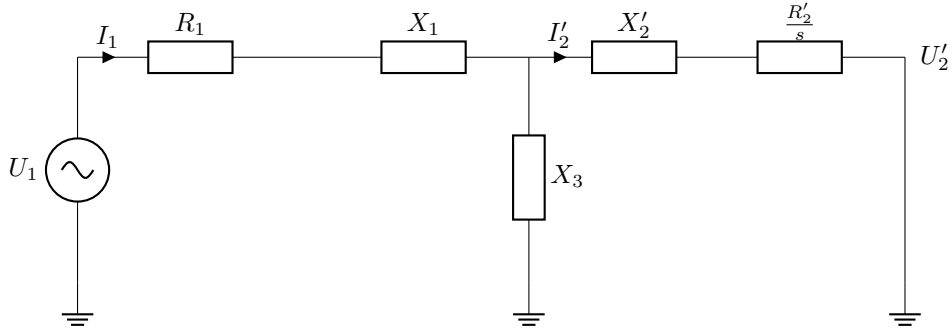

Assuming that the stator current results in a magnetomotive force (MMF) $F_1$ and the rotor current yields an MMF $F_2$, both MMFs rotate synchronously. 
The resulting magnetic field is just the vectorial sum of the two MMFs. 
If the magnetic circuit is assumed to be linear, then the magnetic fields, the fluxes and the electromotive forces (EMFs) from the stator and rotor currents can be summed up. 
If in addition, we add the resistive and inductive leakage voltage drops in the windings, the resulting equivalent circuit for the steady state operating behavior of an asynchronous machine on a symmetrical multiphase network is represented in Figure~\ref{fig:EqCKT}.

The corresponding equations can be derived as
\begin{equation}
    \label{eq:EquCKT}
    \begin{aligned}
        & {{U}_{1}}={{R}_{1}}\cdot {{I}_{1}}+j{{X}_{1 }}\cdot {{I}_{1}}+j{{X}_{3}}\cdot \left( {{I}_{1}}-I'_{2} \right), \\ 
        & U'_{2}=-\frac{R'_{2}}{s}\cdot I'_{2}-jX'_{2 }\cdot I'_{2}+j{{X}_{3}}\cdot \left( {{I}_{1}}-I'_{2} \right),
    \end{aligned}
\end{equation}
where ${U}_{1}$ is the input voltage, ${I}_{1}$ and $I'_{2}$ are the stator and the rotor currents respectively with the prime indicating quantities associated with the stator winding. 
It is important to note that sometimes, depending on the machine type, $U'_{2}$ is set to zero (short circuited) which simplifies the computations. The quantity $s$ is the slip that describes the relation between the rotor speed $n$ and the synchronous speed $\synchronousSpeed$ of the stator flux as

\begin{equation}
    \label{eq:slip}
    s=\frac{\synchronousSpeed-n}{\synchronousSpeed}.
\end{equation}

All the impedance values are chosen to mimic the interactions between the stator and the rotor. 
These values are generally estimated using frequency measurements from the machine \cite{Salon1995}. In view of this, the equivalent circuit model can be considered to be a surrogate model.

Alternatively, the impedance values can be generated based on some simple FEM calculations that fit into the time requirements of the digital twin \cite{MukMSWB20}.

The equations \eqref{eq:EquCKT} are generally solved in phasor form resulting in a set of linear equations  \cite{Nur52,Mel18}. 
The solution gives approximate values for the currents in the stator and the rotor which are then used to compute power and torque. This provides a simple and fast method to compute the currents and can be directly implemented in the digital twin. This also allows for a fast computation of the losses as discussed in the following subsection.

\subsection{Computation of losses and efficiency}
\label{sec:Losses}
One important factor in the study of electrical machines and their efficiency is the total power loss of the machine. 
This represents all the wasted power during the conversion from electrical energy to mechanical energy in the case of motors and the conversion from mechanical to electrical energy in the case of generators. 
Reducing this loss results in an increased efficiency of the machine.
For this reason, monitoring the losses during the lifetime of the machine within the digital twin can help to optimize the design of new machines for a reduced loss and an increased efficiency. In some cases, it is even possible to increase the efficiency online by changing the flux and current to reduce the total loss $\totalLoss$ which is computed as the sum 
\begin{equation}
    \totalLoss = \ironCoreLoss+\frictionLoss+\statorLoss+\rotorLoss+\additionalLoss
\end{equation}
where $\ironCoreLoss$ is the iron core loss (sometimes also called magnetic loss), i.e., the loss that occurs in the stator and rotor cores due to hysteresis and eddy currents. 
The friction losses resulting from the mechanical rotation are summarized in $\frictionLoss$. 
More details on these losses can be found in \cite{YanBKNSPCSE17}. $\statorLoss$ is the stator winding loss and $\rotorLoss$ is the rotor winding loss. These two losses are the result of the inherent resistive properties of windings and conductors. $\additionalLoss$ are all the additional losses that are difficult to quantify and model. Typically these are taken as a global percentage of the total power. More details can be found in~\cite{IqbMR21}. 

For the computation of the losses in the stator and the rotor, the corresponding currents are required as shown in the equations
\begin{equation}
    \label{eq:statorAndRotorLoss}
   \statorLoss =\nPhases R_1 I_1^2, \quad \rotorLoss =\nPhases R'_2 (I'_2)^{2},
\end{equation}
where $\nPhases$ indicates the number of phases of the system.
The currents are typically computed directly based on the equivalent circuit described in the previous subsection. It is important to note that the material properties change with increased temperatures, hence the resistance changes with the temperature
of the conductor. Typically, these changes are neglected when using this type of simulation models. Instead, a typical operating temperature is used to compute for example $R_1$ and $R'_2$.

On the other hand, using computational models based on discretizing the PDEs, as discussed in section~\ref{sec:electromagnetics_model}, may be computationally prohibitively expensive. If in addition one wants to do parameter optimization, then repeated simulations for different parameter configurations are required, rendering the determination of a solution in real-time within the digital twin impossible. 
One approach for avoiding such difficulties is to construct a parametric model that could be reduced using model order reduction and used in the optimization step. In addition, depending on the model reduction technique used, one may also obtain error bounds. Alternatively, one could use the equivalent circuit model, however, the results may not be as accurate and are typically without  error estimates.

\section{Mechanical simulation}

An important step of the design process of electrical machines is the determination of maximal mechanical stresses, since they limit (among other effects) the power of an electrical machine, cf.~\cite{GerZP21}.
Furthermore, vibration and noise analysis are typical mechanical calculations performed for an electrical machine in order to reduce acoustic noise emissions, cf.~\cite{BoeSKD10,GieWL06,JavLCL95,PelLF12,RamHKH96,XuHC18}.
Both aspects are discussed in section~\ref{sec:vibrations}, whereas we address the relationship between the electromagnetic torque and the resulting rotational speed in section~\ref{sec:rotationSpeed} via Newton's second law for rotation.
\subsection{Analysis of displacements and vibrations}
\label{sec:vibrations}

In kinematics one is typically interested in deformations and motions of a physical body with respect to a reference configuration. 
A configuration or deformation is described by a mapping from the reference configuration $\Refconfig$ to $\R^3$ and a motion is a time-dependent family of deformations, see for instance \cite{MarH83} for more details.
The simplest examples for motions are rigid body motions such as translations or rotations.
In general, the time evolution of the motion is governed by the balance of momentum which, for a solid continuous body, can be written  in Lagrangian coordinates as
\begin{equation}
    \label{eq:balanceOfMomentum}
    \rhoref(\lag)\frac{\partial^2 \motion}{\partial t^2}(t,\lag) = \nabla\cdot\fPiola(t,\lag)+\rhoref(\lag)\forceField(t,\lag),
\end{equation}
where $\lag\in\Refconfig$ is an arbitrary point in the reference configuration, $\motion\colon \R_{\ge 0}\times \Refconfig\to \R^3$ is the motion of the considered body, $\fPiola\colon \R_{\ge 0}\times \Refconfig\to \R^{3,3}$ denotes the first Piola--Kirchhoff stress tensor, $\rhoref\colon \Refconfig\to \R_{>0}$ is the density in the reference configuration, and $\forceField\colon \R_{\ge 0}\times \Refconfig\to \R^3$ is an external body force, cf.~\cite[ch.~1--2]{MarH83}.
Here, the divergence of $\fPiola$ is defined via

\begin{equation*}
	[\nabla\cdot\fPiola]_i \vcentcolon= \sum_{j=1}^3 \frac{\partial \fPiola_{ij}}{\partial \lagComponent_j}\quad \text{for }i=1,\ldots,3.
\end{equation*}

The unknown of \eqref{eq:balanceOfMomentum} is the motion $\motion$, whereas the density $\rhoref$ and the external force $\forceField$ are assumed to be given.
For instance, in the context of electrical machines, the latter term usually includes electromagnetic forces, see for instance \cite{XuHC18}.
The stress tensor $\fPiola$ is specified by means of constitutive laws.
For instance, a homogeneous and elastic material can be described by a constitutive law of the form
\begin{equation*}
    \fPiola(t,\lag) = \fPiolaConsti\left(\frac{\partial \motion}{\partial \lag}(t,\lag)\right),
\end{equation*}
which is characterized by the constitutive function $\fPiolaConsti\colon \R^{3,3}\to \R^{3,3}$, cf.~\cite[p.~8]{MarH83}.
The tensor $\frac{\partial\motion}{\partial\lag}$ is called the deformation tensor and denoted by $\deformationtensor$ in the following.
If one considers viscous dissipation, then the constitutive law of a homogeneous material takes the form
\begin{equation}
    \label{eq:generalViscoelasticLaw}
    \fPiola(t,\lag) = \fPiolaConsti\left(\deformationtensor(t,\lag),\frac{\partial \deformationtensor}{\partial t}(t,\lag)\right),
\end{equation}
with constitutive function $\fPiolaConsti\colon \R^{3,3}\times \R^{3,3}\to \R^{3,3}$, cf.~\cite[sec.~13.9]{Ant05b}.
Finally, \eqref{eq:balanceOfMomentum} is closed by initial conditions
\begin{equation*}
    \motion(0,\lag) = \motion_0(\lag),\quad \frac{\partial \motion}{\partial t}(0,\lag) = \velocity_0(\lag)
\end{equation*}
with initial deformation $\motion_0$ and velocity $\velocity_0$, as well as by appropriate boundary conditions.
Typical boundary conditions are displacement boundary conditions of the form
\begin{equation*}
    \motion(t,\lag) = \boundarymotion(t,\lag)
\end{equation*}
for $\lag\in\Gamma_1$ with prescribed boundary deformation $\boundarymotion$ and traction boundary conditions of the form
\begin{equation}
    \label{eq:boundaryTraction}
    \fPiola(t,\lag)\outwardnormal(\lag) = \boundarytraction(t,\lag)
\end{equation}
for $\lag\in\Gamma_2$ with $\Gamma_1\cup\Gamma_2=\partial\Refconfig$, $\Gamma_1\cap\Gamma_2=\emptyset$, outer unit-normal vector field $\outwardnormal$ on the boundary $\partial\Refconfig$, and a prescribed boundary traction $\boundarytraction$.
For some examples of such boundary conditions, we refer to \cite[p.~11~ff.]{MarH83}.

\begin{remark} {\rm
    In this work, we only consider constitutive laws of the form \eqref{eq:generalViscoelasticLaw}.
    However, we emphasize that in general $\fPiolaConsti$ may not only depend on the values of $\deformationtensor=\frac{\partial \motion}{\partial \lag}$ and $\frac{\partial \deformationtensor}{\partial t}=\frac{\partial^2 \motion}{\partial t\partial \lag}$, but on $\motion$ as a function.
    Thus, also integrals or higher-order derivatives of $\motion$ may occur in the constitutive laws.
    Furthermore, the stress may not only depend on the motion $\motion$, but also on the temperature or the time, and for non-homogeneous materials even on the position $\lag$, cf.~\cite[ch.~3]{MarH83}.
    In the context of electrical machines the temperature dependence of the stress has for example been considered in \cite{SatBKVS16,SilS18}.}
\end{remark}

In the following, we consider the special case 
\begin{equation}
    \label{eq:KelvinVoigt}
    \fPiolaConsti(F_1,F_2) \vcentcolon= \frac12F_1\left(\elasticity(F_1^\top F_1-I_3)+\viscosity(F_2^\top F_1+F_1^\top F_2)\right)
\end{equation}
of the constitutive law \eqref{eq:generalViscoelasticLaw}, which corresponds to a Kelvin--Voigt model for the material, see for instance \cite{GioM21,LinLT13}.
Here, $\elasticity,\viscosity\in\R^{3,3,3,3}$ are the so-called instantaneous or Hookean elasticity tensor and the viscosity tensor, respectively.
Furthermore, we restrict ourselves to the special case of a homogeneous and isotropic material such that $\elasticity$ and $\viscosity$ can be specified as 
\begin{equation}
    \label{eq:fourthOrderTensors}
    \elasticity = \firstLame I_3\otimes I_3+2\secondLame\fourthorderidentity\quad \text{and}\quad \viscosity = \left(\bulkvisc-\frac23\shearvisc\right)I_3\otimes I_3+2\shearvisc \fourthorderidentity,
\end{equation}
cf.~\cite{LinLT13}.
Here $\firstLame,\secondLame\in\R_{\ge 0}$ are the so-called Lam\'e parameters, $\bulkvisc,\shearvisc\in\R_{\ge 0}$ are the bulk and the shear viscosity, respectively, $I_3\in\R^{3,3}$ is the $3\times 3$ identity matrix, $\otimes$ denotes the Kronecker product, and $\fourthorderidentity\in\R^{3,3,3,3}$ is the fourth-order identity tensor, i.e., $\fourthorderidentity$ is the unique fourth-order tensor satisfying $\fourthorderidentity A=A$ for all $A\in\R^{3,3}$.

For practical calculations, often linearized equations of motion are considered by employing an assumption of small displacements or strains.
In the context of static components like the stator in an electric machine, the small displacements are considered with respect to the reference configuration.
On the other hand, when considering rotating components like the rotor of an electric machine, the small displacements are considered with respect to a rigid body rotation.
Both cases are addressed separately in sections~\ref{sec:nonRotatingComponents} and \ref{sec:rotatingComponents}.
Furthermore, in section~\ref{sec:MORmechanics} we address previous studies applying model order reduction for mechanical simulations in the context of electrical machines.

\subsubsection{Dynamics of non-rotating components}
\label{sec:nonRotatingComponents}

The constitutive law \eqref{eq:KelvinVoigt} especially satisfies $\fPiolaConsti(I_3,0)=0$, i.e., the trivial motion $\motion(t,\lag)=\lag$ has zero stress.
Consequently, a particular solution of \eqref{eq:balanceOfMomentum} is given by $\motion(t,\lag)=\lag$ which corresponds to the external force $\forceField=0$.
In many works concerned with a mechanical analysis of the static components of an electrical machine, the displacements are assumed to be small and, thus, linearized equations of motions are considered, see e.g.~\cite{SatBKVS16,Wer17}.
More precisely, the governing equations \eqref{eq:balanceOfMomentum} are linearized around the solution $\motion(t,\lag)=\lag$.
This is done by substituting the ansatz $\motion(t,\lag) = \lag+\epsilon \smalldisp(t,\lag)$, $\forceField(t,\lag) = \epsilon\smallforce(t,\lag)$, differentiating the resulting equation with respect to $\epsilon$, and setting $\epsilon=0$.
The resulting linearized equation reads
\begin{equation}
    \label{eq:linearizedMomentumBalanceForNonrotatingComponents}
    \rhoref(\lag)\frac{\partial^2\smalldisp}{\partial t^2}(t,\lag) = \nabla\cdot\left(\elasticityHat \frac{\partial\smalldisp}{\partial\lag}+\viscosityHat\frac{\partial^2\smalldisp}{\partial t\partial\lag}\right)(t,\lag)+\rhoref(\lag)\smallforce(t,\lag),
\end{equation}
where $\elasticityHat$ and $\viscosityHat$ are given by
\begin{equation}
    \label{eq:newTensors}
    \elasticityHat = \firstLame I_3\otimes I_3+\secondLame(\fourthorderidentity+\transposer)\quad \text{and}\quad \viscosityHat = \left(\bulkvisc-\frac23\shearvisc\right)I_3\otimes I_3+\eta(\fourthorderidentity+\transposer)
\end{equation}
and $\transposer\in\R^{3,3,3,3}$ denotes the unique fourth-order tensor satisfying $\transposer A=A^\top$ for all $A\in\R^{3,3}$.
This linearization procedure is for instance illustrated in \cite[p.9~f.]{MarH83} for the case without dissipation, i.e., $\bulkvisc=\shearvisc=0$.
The major model assumptions leading to \eqref{eq:linearizedMomentumBalanceForNonrotatingComponents} may be summarized as follows:
\begin{itemize}
    \item The material may be described by a constitutive law of the form \eqref{eq:KelvinVoigt}  with $\elasticity$ and $\viscosity$ as in \eqref{eq:fourthOrderTensors}.
    \item The displacement $\motion-\lag$ and the external force $\forceField$ are assumed to be small.
\end{itemize}

After semi-discretization in space by a Galerkin projection based on finite element basis functions, the semi-discretized governing equations are of the form
\begin{equation}
    \label{eq:semidiscreteMomentumBalance}
    M\ddot{\dispVec}(t)+D\dot{\dispVec}(t)+K\dispVec(t) = \discForceVec(t),
\end{equation}
where $\dispVec(t)\in\R^N$ denotes the vector of displacements at time $t$, $M\in\R^{N,N}$ is the mass matrix, $D\in\R^{N,N}$ the damping matrix, $K\in\R^{N,N}$ the stiffness matrix, and $\discForceVec(t)\in\R^N$ the vector of external forces at time $t$, cf.~\cite{MesC02}.

Instead of deriving \eqref{eq:semidiscreteMomentumBalance} from a semi-discretization of a partial differential equation, it is also possible to arrive at a system like \eqref{eq:semidiscreteMomentumBalance} by modeling the flexible body by a mass-spring-damper system, see for instance \cite[Ch.~9]{CloP03}.

Sometimes, researchers also follow a hybrid approach, where the mass and the stiffness matrix are derived from a finite element discretization and the damping matrix is for instance obtained by a surrogate data based ansatz, or the assumption of Rayleigh damping, i.e., as a linear combination of the mass and the stiffness matrix, cf.~\cite[sec.~9.3.3]{Bat14}.
The coefficients of this linear combination may for example be tuned based on data from a digital twin and updated during the lifetime of the machine, see for instance \cite{KapPW21}.
It is also common to completely neglect damping effects and consider the undamped system
\begin{equation*}
    M\ddot{\dispVec}(t)+K\dispVec(t) = \discForceVec(t),
\end{equation*}
see e.g.~\cite{HenSHS92} or \cite[sec.~9.3.2]{Bat14}.

\subsubsection{Dynamics of rotating components}
\label{sec:rotatingComponents}

In the case of rotating components like the rotor of an electrical machine, the assumption of small displacements as discussed in the last subsection is often not valid, since the rotation leads to large displacements.
Nevertheless, in a co-rotating frame, the displacements may still be considered to be small.
In the following, we assume the case of a rigid-body rotation in the $\lagComponent_1\lagComponent_2$-plane with constant rotation velocity $\rotVel$.
This rotation may be described by the rotation matrix
\begin{equation}
    \rotMat(t) =
    \begin{bmatrix}
        \cos(\rotVel t)     & -\sin(\rotVel t)  & 0\\
        \sin(\rotVel t)     & \cos(\rotVel t)   & 0\\
        0                   & 0                 & 1
    \end{bmatrix}
    .
\end{equation}
Especially, we note that the constitutive law \eqref{eq:KelvinVoigt} satisfies $\fPiolaConsti(\rotMat(t),\dot{\rotMat}(t))=0$ for all $t\ge 0$, i.e., the rigid body rotation described by $\motion(t,\lag) = \rotMat(t)\lag$ is stress-free.
This follows from \eqref{eq:KelvinVoigt} and the facts that $\rotMat(t)$ is orthogonal and
\begin{equation*}
    \dot{\rotMat}(t)^\top\rotMat(t) = \rotVel 
    \underbrace{
    \begin{bmatrix}
        0   & 1 & 0\\
        -1  & 0 & 0\\
        0   & 0 & 0
    \end{bmatrix}
    }_{=\vcentcolon J}
\end{equation*}
is skew-symmetric for any $t\ge 0$.
Consequently, a particular solution of \eqref{eq:balanceOfMomentum} is given by $\motion(t,\lag) = \rotMat(t)\lag$ which corresponds to the external force $\forceField(t,\lag) = \ddot{\rotMat}(t)\lag$.
Substituting the ansatz $\motion(t,\lag) = \rotMat(t)(\lag+\epsilon \smalldisp(t,\lag))$, $\forceField(t,\lag) = \ddot{\rotMat}(t)\lag+\epsilon\smallforce(t,\lag)$, differentiating the resulting equation with respect to $\epsilon$, setting $\epsilon=0$, and multiplying from the left by $\rotMat(t)^\top$ leads to the linearized equation
\begin{equation}
    \label{eq:linearizedMomentumBalanceForRotatingComponents}
    \rhoref(\lag)\left(\frac{\partial^2\smalldisp}{\partial t^2}(t,\lag)+2\rotVel J
  \frac{\partial\smalldisp}{\partial t}(t,\lag)-\rotVel^2\smalldisp(t,\lag)\right) = \nabla\cdot\left(\elasticityHat \frac{\partial\smalldisp}{\partial\lag}+\viscosityHat\frac{\partial^2 \smalldisp}{\partial t\partial\lag}\right)(t,\lag)+\rhoref(\lag)\rotMat(t)^\top\smallforce(t,\lag)
\end{equation}
with $\elasticityHat$ and $\viscosityHat$ as defined in \eqref{eq:newTensors}.
A similar linearization approach has been considered in \cite{GraMM07} in the context of linear elasticity.
Similarly as in the previous subsection, the major model assumptions leading to \eqref{eq:linearizedMomentumBalanceForRotatingComponents} may be summarized as follows:
\begin{itemize}
    \item The material may be described by a constitutive law of the form \eqref{eq:KelvinVoigt}  with $\elasticity$ and $\viscosity$ as in \eqref{eq:fourthOrderTensors}.
    \item The relative displacement $\motion-\rotMat\lag$ and the relative external force $\forceField-\ddot{\rotMat}\lag$ are assumed to be small.
\end{itemize}

Semi-discretization in space via a Galerkin finite element method leads to the semi-discretized system
\begin{equation}
    \label{eq:rotorDynamics}
    M\ddot{\dispVec}(t)+(D+2\rotVel \gyroscopicMatrix)\dot{\dispVec}(t)+(K-\rotVel^2 Z)\dispVec(t) = \discForceVec(t),
\end{equation}
where $\dispVec(t)\in\R^N$ is the displacement vector in the co-rotating coordinate system at time $t$, $\gyroscopicMatrix\in\R^{N,N}$ the gyroscopic matrix, $Z\in\R^{N,N}$ the centrifugal stiffness matrix, and $\discForceVec(t)\in\R^N$ the vector of external forces in the co-rotating coordinate system at time $t$, cf.~\cite{Kir16} for an alternative derivation via an Euler--Lagrange approach. 
Also, a lumped-parameter modeling approach may lead to a similar equation system as the one in \eqref{eq:rotorDynamics}, cf.~\cite[ch.~4]{Gen05}. 
Alternatively, one can use a data-driven method to estimate the values of the parameters in the lumped-parameter model as in \cite[ch.~3]{PoG19}.

An effect which is not captured by the linearized theory leading to \eqref{eq:rotorDynamics} is the so-called centrifugal stiffening effect, see for instance \cite{SimV87}.
This effect leads to a higher bending stiffness of a rotating beam in comparison to a non-rotating beam, cf.~\cite{BerS02}.
To take this effect into account, usually a term of the form $\rotVel^2\KG \dispVec(t)$ with $\KG\in\R^{N,N}$ is added to the left-hand side of \eqref{eq:rotorDynamics}, see \cite{Kir16,SimV87}.

In \cite{ClaG17}, the authors investigated electromagnetic noise within an electrical machine by considering a model of the form \eqref{eq:rotorDynamics} for the rotor, with an additional geometric stiffness matrix and without viscous damping effects, i.e., they neglected the $D$ term.

In \cite{ArkNS10}, the authors investigate the stability of a whirling rotor by a Jeffcott rotor model with one complex-valued degree of freedom or, equivalently, with two real-valued degrees of freedom of the form
\begin{equation*}
    m \ddot{\dispVec}(t)+d\dot{\dispVec}(t)+k\dispVec(t) = \discForceVec(t,\dispVec(t)),
\end{equation*}
where $m$ is the rotor mass, $d$ the damping coefficient, $k$ the stiffness constant, $\discForceVec$ the time- and state-dependent external forces, and $\dispVec(t)\in\R^2$ the coordinate vector of the displacement at time $t$, see also \cite[sec.~2.4]{Gen05}.
This model does neither take structural damping into account nor gyroscopic forces.
The force vector $\discForceVec$ includes forces caused by the eccentricity of the rotor and electromagnetic forces.
The latter ones are dependent on the displacements and in \cite{ArkNS10} the authors calculate these forces by adding two further complex-valued differential equations to the system.
The resulting overall differential equation system is linear and the stability is investigated by computing the poles of the corresponding transfer function.

\subsubsection{Model order reduction for the mechanical simulation}
\label{sec:MORmechanics}

Especially, when using full 3D models for the mechanical behavior of the rotor and the stator, the simulation times are very high and prevent any real-time applications.
As pointed out in the last two sections, then model order reduction is a useful tool to reduce the computational burden.
In \cite{SchBB14}, the authors use a Krylov subspace method to obtain a reduced-order model (ROM) for simulating the stator vibrations within an electric motor.
To this end, the authors consider the mapping from an input vector which corresponds to external forces to an output vector which coincides with the displacement vector. The Arnoldi algorithm, see e.g.~\cite{LehS98}, is then applied to the transfer function of the undamped system to determine a suitable Krylov subspace and a corresponding projection matrix, which is afterwards used to obtain the reduced-order model via projection of the original damped system.
The article  also compares this MOR approach with the classical mode superposition method, where the projection matrix is based on the eigenmodes, and observes a superior performance of the reduced-order model obtained by the Krylov technique.
Similarly, in \cite{EseGMMM19} a Krylov subspace method has been proposed for constructing a reduced-order model for investigating the noise radiated from the stator.
In contrast to \cite{SchBB14}, they consider only a few pre-defined displacements as output vector and they apply the Krylov technique directly to the damped system by employing a second-order Arnoldi algorithm.
A model reduction scheme for the rotor dynamics has been proposed in \cite{Kot19} based on the POD method, where the POD basis captures the dynamics for a certain range of the rotor speed $\rotVel$.
For further model reduction approaches for mechanical systems in general, we refer to the references in \cite{BedBDRSVW19,LuZZGJZFY21}.

The simulation of the mechanical full-order model (FOM) is often based on commercial software.
In particular, \textsc{Ansys Mechanical} is frequently used, cf.~\cite{BoeSKD10,EseGMMM19,Kot19}, but also others such as \textsc{NASTRAN}, e.g.~in \cite{PelLF12}, or \textsc{Abaqus FEA}, cf.~\cite{ClaG17}.
In \cite{SilS18}, where also the coupling to the thermal field is considered, the authors use the software \textsc{COMSOL Multiphysics} to investigate the stresses caused by thermal effects.
In some works, open source software packages are employed, as for instance \textsc{Elmer} which is used in \cite{SatBKVS16}.
Some of the mentioned commercial software packages have also built-in model reduction tools.
However, to allow for more flexibility in the choice of the MOR scheme and to be able to set up a model hierarchy including error bounds, it may be more effective to export the coefficient matrices from the software and, afterwards, perform MOR based on these extracted matrices.
For \textsc{Ansys}, this extraction and subsequent MOR has been, for example, demonstrated in \cite{RudK06} and for \textsc{Abaqus} in \cite{Her08}.
Furthermore, the software \textsc{LiveLink} allows to export the system matrices from \textsc{COMSOL Multiphysics} to \textsc{MATLAB}, which has been exploited in \cite{HabDHCK21} for model order reduction.

\subsection{Calculation of the rotational speed}
\label{sec:rotationSpeed}

In section~\ref{sec:rotatingComponents} we have considered the case of a rotor with constant rotational speed.
Especially, in the start-up and in the shut-down phase, as well as when passing from one operation speed to another, the rotation speed is not constant, but instead a result of an equilibrium of torques.
A simple model for simulating the rotational speed of the rotor shaft based on the electromagnetic torque and the load torque is given by
\begin{equation}
    \label{eq:torqueBalance}
    \torqueofinertia\dot{\rotVel}(t) = -\frictionCoeff\rotVel(t)+\Te(t)-\TL(t),
\end{equation}
where $\rotVel$ denotes the motor speed, $\torqueofinertia$ the torque of inertia, $\frictionCoeff$ the friction coefficient, $\Te$ the electromechanical torque, and $\TL$ the load torque, see for instance \cite{DamHK21}.
Neglecting the friction in \eqref{eq:torqueBalance} leads to the simplified undamped model
\begin{equation}
    \label{eq:torqueBalance_undamped}
    \torqueofinertia\dot{\rotVel}(t) = \Te(t)-\TL(t),
\end{equation}
cf.~\cite[sec.~2.20.3]{HraRM20}.

\section{Thermal simulation}
\label{sec:thermalSimulation}

In this section the simulation of the temperature distribution within an electrical machine is discussed. There are several reasons which lead to a need for temperature simulations in electrical machines.
For instance, high temperatures usually lead to high resistances in copper which result in turn in an increase of copper losses and, thus, it has also an effect on the overall efficiency.
Also ageing and damage of the windings can be facilitated by an excessive thermal load, cf.~\cite{YanBKNSPCSE17}.
Thus, a thermal analysis by means of numerical simulations is helpful for a suitable design of the cooling system which prevents excessively high temperatures in the machine. 
A higher temperature than expected could also be  a sign of a fault in the system and thus tracking the temperature trend in the machine can help in predictive maintenance.

The most relevant approaches for thermal simulation can roughly be subdivided into two classes, see for instance \cite{Gra92,YanBKNSPCSE17}.
Approaches which are based on the numerical discretization of the heat conduction equation are discussed in section~\ref{sec:thermal_discretization}, whereas so-called \emph{thermal equivalent circuits}, which are based on lumped-parameter representations, are addressed in section~\ref{sec:thermalNetworks}.

\subsection{Energy balance and numerical discretization schemes}
\label{sec:thermal_discretization}

The time evolution of the temperature field is primarily governed by the balance of energy, which reads in general
\begin{equation}
    \label{eq:generalEnergyBalance}
    \rhoref(\lag)\frac{\partial\intEnergy}{\partial t}(t,\lag) = -\nabla\cdot \heatFlux(t,\lag)+\frac12\ipF{\sPiola(t,\lag),\frac{\partial}{\partial t}(\deformationtensor^\top\deformationtensor)(t,\lag)}+\rhoref(\lag)\heatsource(t,\lag).
\end{equation}
Here, $\intEnergy\colon\R_{\ge 0}\times \Refconfig\to \R$ denotes the specific internal energy, $\heatFlux\colon\R_{\ge 0}\times \Refconfig\to \R^3$ the heat flux vector, $\sPiola\vcentcolon= \deformationtensor^{-1}\fPiola$ the second Piola--Kirchhoff stress tensor, $\deformationtensor$ the deformation tensor introduced in the previous section, $\ipF{\cdot,\cdot}$ the Frobenius inner product, and $\heatsource\colon\R_{\ge 0}\times \Refconfig\to \R$ the specific heat supply, cf.~\cite[p.~145]{MarH83}.
Moreover, one can show the identity 
\begin{equation}
    \label{eq:ipF_stress_deformationRate}
    \frac12\ipF{\sPiola,\frac{\partial}{\partial t}(\deformationtensor^\top\deformationtensor)}=\ipF{\fPiola,\frac{\partial F}{\partial t}},
\end{equation}
where $\fPiola$ denotes the first Piola--Kirchhoff stress tensor as introduced in the previous section, cf.~\cite[p.~144]{MarH83}.
Furthermore, the specific internal energy $\intEnergy$ may be expressed in terms of the Helmholtz free energy $\freeenergy$, the temperature $\temperature$, and the specific entropy $\entropy$ via 
\begin{equation}
    \label{eq:freeEnergy}
    \intEnergy = \freeenergy+\temperature\entropy.
\end{equation}
The system is then closed via constitutive equations for $\freeenergy$, $\entropy$, and $\heatFlux$, whereas the temperature $\temperature$ is the unknown of the energy equation.
To be consistent with the considerations from section~\ref{sec:vibrations}, we consider again a Kelvin--Voigt material which results in a constitutive law $\freeenergy = \freeenergyConsti(\temperature,\deformationtensor)$ of the form
\begin{equation}
    \label{eq:freeenergyConsti}
    \freeenergyConsti(\temperature,\deformationtensor) = \freeenergy_0(\temperature)+\frac1{8\rhoref}\ipF{\elasticity(\deformationtensor^\top\deformationtensor-I_3),\deformationtensor^\top\deformationtensor-I_3},
\end{equation}
where $\freeenergy_0$ is a given function of the temperature representing the free energy in a deformation-free state and $\elasticity$ is the elasticity tensor as defined in \eqref{eq:fourthOrderTensors}, cf.~\cite[sec.~6.3]{GioM21}.
The specific entropy can be derived from the free energy via the relation
\begin{equation}
    \label{eq:entropy}
    \entropy = -\frac{\partial\freeenergyConsti}{\partial \temperature} = -\frac{\partial\freeenergy_0}{\partial \temperature},
\end{equation}
cf.~\cite[sec.~3]{GioM21}.
Furthermore, for the heat flux $\heatFlux$ we use Fourier's law
\begin{equation}
    \label{eq:FouriersLaw}
    \heatFlux = -\thermalConductivity\nabla\temperature,
\end{equation}
where $\thermalConductivity\in\R^{3,3}$ is the thermal conductivity matrix, cf.~\cite[sec.~7.1.2]{Dim11}.
Finally, by substituting \eqref{eq:freeEnergy}, $\freeenergy = \freeenergyConsti(\temperature,\deformationtensor)$, \eqref{eq:entropy}, \eqref{eq:FouriersLaw}, \eqref{eq:ipF_stress_deformationRate}, and $\fPiola = \fPiolaConsti(\deformationtensor,\frac{\partial \deformationtensor}{\partial t})$ into \eqref{eq:generalEnergyBalance} and by using the constitutive equations \eqref{eq:freeenergyConsti} and \eqref{eq:KelvinVoigt}, we arrive at the heat conduction equation
\begin{equation}
    \label{eq:heatConduction}
    \rhoref(\lag)\specificHeat(\temperature(t,\lag))\frac{\partial\temperature}{\partial t}(t,\lag) = \nabla\cdot(\thermalConductivity\nabla \temperature)(t,\lag)+\ipF{\viscosityHat \left(\deformationtensor(t,\lag)^\top\frac{\partial \deformationtensor}{\partial t}(t,\lag)\right),\deformationtensor(t,\lag)^\top\frac{\partial \deformationtensor}{\partial t}(t,\lag)}+\rhoref(\lag)\heatsource(t,\lag),
\end{equation}
where
\begin{equation}
    \label{eq:specificHeat}
    \specificHeat(\temperature)\vcentcolon=-\temperature\freeenergy_0^{\prime\prime}(\temperature)
\end{equation}
is the specific heat at constant strain, cf.~\cite[eq.~(8)]{BerH93}, and $\viscosityHat$ is as defined in \eqref{eq:newTensors}.
The model assumptions leading to \eqref{eq:heatConduction} are essentially the constitutive laws \eqref{eq:KelvinVoigt} and \eqref{eq:freeenergyConsti}--\eqref{eq:FouriersLaw} with $\elasticity$ and $\viscosity$ as in \eqref{eq:fourthOrderTensors}.

The heat source due to mechanical friction enters \eqref{eq:heatConduction} via the second term on the right-hand side and other heat sources enter the equation via the $\heatsource$ term, which may in general also depend on the temperature.
For instance, when an electric current flows through the material, this is usually accompanied by Joule heating, which corresponds to a heat source of the form
\begin{equation*}
    \rhoref\JouleHeat = \currentDensityLag\cdot\electricFieldLag,
\end{equation*}
where $\currentDensityLag$ denotes the current density and $\electricFieldLag$ the electric field, cf.~section~\ref{sec:electricalSim}.
In many cases, a linear relationship between $\currentDensityLag$ and $\electricFieldLag$ may be assumed leading to $\currentDensityLag = \electricConductivity\electricFieldLag$, where $\electricConductivity$ denotes the electrical conductivity, see for instance \cite[\S~20]{LanL60}.
Another origin of heat sources is a time-varying magnetic field which leads e.g.~to hysteresis losses, cf.~\cite[sec.~6.2]{Sud14}, \cite[sec.~10.3.1]{EriM20}, \cite[sec.~10.3]{Fer16}, and eddy current losses, cf.~\cite[sec.~6.1]{Sud14}.
For a more extensive discussion of the different origins of losses within electrical machines, we refer to \cite[sec.~4]{YanBKNSPCSE17}.

The partial differential equation \eqref{eq:heatConduction} is closed by an initial condition of the form
\begin{equation*}
    \temperature(0,\lag) = \temperature_0(\lag)\quad \text{for } \lag\in\Refconfig
\end{equation*}
with initial temperature profile $\temperature_0$ and by boundary conditions of the form
\begin{align}
    \temperature(t,\lag) &= \boundarytemperature(t,\lag)\quad \text{for } \lag\in\Gamma_1,\\
    \label{eq:temperatureNeumannBC}
    \heatFlux(t,\lag)^\top\outwardnormal(\lag) &= \boundaryheatflow(t,\lag)\quad \text{for }\lag\in\Gamma_2,\\
    \label{eq:Robin}
    w_1\heatFlux(t,\lag)^\top\outwardnormal(\lag)+w_2\temperature(t,\lag) &= \Robinvalue(t,\lag)\quad \text{for }\lag\in\Gamma_3,
\end{align}
with $\Gamma_1\cup\Gamma_2\cup\Gamma_3=\partial\Refconfig$, $\Gamma_i\cap\Gamma_j=\emptyset$ for $i\ne j$, prescribed boundary temperature $\boundarytemperature$, prescribed boundary heat flow $\boundaryheatflow$, and outer unit-normal vector field $\outwardnormal$ on the boundary $\partial\Refconfig$, cf.~\cite[p.~211]{MarH83}.
Furthermore, the Robin boundary condition \eqref{eq:Robin} is determined by the prescribed function $\Robinvalue$ and by the weighting parameters $w_1,w_2\in\R$.
Typical in the context of electrical machines are Neumann boundary conditions of the form \eqref{eq:temperatureNeumannBC} and Robin boundary conditions of the form \eqref{eq:Robin}, see for instance \cite{ChaZTF00,FanZQLLH19}.
One approach for simulating the temperature distribution within an electrical machine is based on discretizing the heat conduction equation.
The space discretization is usually performed via the finite element method, cf.~\cite{FanZQLLH19,ChaZTF00}.
In \cite{ZhuCC18} the authors use a coupled FEM-circuit model for describing the heat transfer within an electrical machine, i.e., the finite element method is used for some components of the electrical machine and a thermal equivalent circuit, cf.~section~\ref{sec:thermalNetworks}, for other components.

Since the heat transfer from the motor components to the surrounding coolant depends also on the fluid flow of the coolant, numerical simulation of the heat transfer is often also coupled to a simulation of the coolant's fluid dynamics.
For instance, in \cite{KraHHLK09} the authors use a computational fluid dynamics (CFD) approach for simulating the coupled heat and fluid flow under steady state conditions. 
The coupling of steady-state heat transfer, fluid flow, and electromagnetics was addressed in \cite{ZhaRHYZY12} using the finite element method.

\subsection{Thermal equivalent circuits}
\label{sec:thermalNetworks}

For practical temperature calculations for electrical machines, lumped-parameter thermal networks are often used instead of an explicit spatial discretization of the heat conduction equation, see for instance \cite{BogCSSMM09,CamHM16,DurF04,GuoHGZ16,LieS69,MelRT91,PreTP69} and the references therein.
The main idea of the thermal equivalent circuit modeling approach is to replace the spatially distributed heat conduction equation by a discrete network of temperature nodes. 
This is done by decomposing the spatial domain into several subdomains and assuming a homogeneous temperature for each of these subdomains.
The resulting thermal equivalent circuit model shares analogies with electrical resistor-capacitor circuits, where heat transfer between the subdomains
is modeled via a conductor including a thermal resistor and the heat storage within each subdomain is described by a thermal capacitance.
Furthermore, there is a heat source term in each subdomain which corresponds to a current source in the electrical circuit analogy, see for instance \cite{DjaT17}.
A detailed derivation of thermal equivalent circuits based on the heat conduction equation is for instance presented in \cite[Ch.~10]{Sud14}.

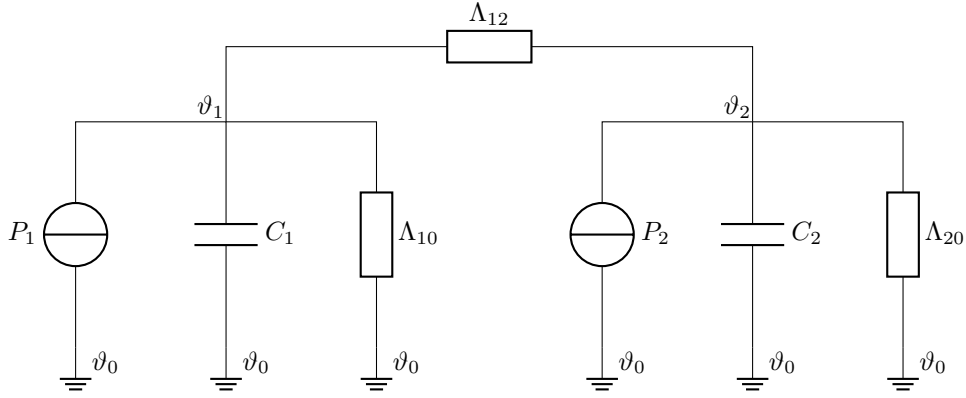
\begin{figure}
	\centering
	\begin{circuitikz}[european]
		\draw node[ground]{} (0,0) to[isource, l=$P_1$] (0,3) -- (2,3) to[capacitor=$C_1$] (2,0) node[ground]{};
		\draw (2,3) -- (4,3) to[R=$\Lambda_{10}$] (4,0) node[ground]{};
		\draw (9,3) -- (7,3) to[isource, l=$P_2$] (7,0) node[ground]{} ;
		\draw (9,3) to[capacitor=$C_2$] (9,0) node[ground]{};
		\draw (9,3) -- (11,3) to[R=$\Lambda_{20}$] (11,0) node[ground]{};
		\draw (2,3) -- (2,4) to[R=$\Lambda_{12}$] (9,4)--(9,3) ;
		\node (a) at (1.8,3.2) {$\temperature_1$};
		\node (b) at (8.8,3.2) {$\temperature_2$};
		\node (c) at (0.4,-0.2) {$\temperature_0$};
		\node (d) at (2.4,-0.2) {$\temperature_0$};
		\node (e) at (4.4,-0.2) {$\temperature_0$};
		\node (f) at (7.4,-0.2) {$\temperature_0$};
		\node (g) at (9.4,-0.2) {$\temperature_0$};
		\node (h) at (11.4,-0.2) {$\temperature_0$};
	\end{circuitikz}
	\caption{Example of a thermal equivalent circuit with two subdomains}
	\label{fig:twoBodyCircuit}
\end{figure}

An example for a thermal equivalent circuit with two subdomains is depicted in Figure~\ref{fig:twoBodyCircuit}.
Each subdomain consists of a heat source term $P_i$ (current source in circuit analogy), a thermal capacitance $C_i$ (a capacitor in circuit analogy), and a thermal resistance $\Lambda_{i0}$ (resistor in circuit analogy) for the heat transfer to the environment or to a cooling medium with temperature $\temperature_0$, which is assumed to be constant in the following.
The temperature $\temperature_i$ of a subdomain corresponds in the circuit analogy to the electric potential at the node where the three elements of the subdomain are connected. 
Furthermore, the heat transfer between different subdomains is also described by a thermal resistance, which is $\Lambda_{12}$ in the example from Figure~\ref{fig:twoBodyCircuit}.

The governing equations of a thermal equivalent circuit are derived by applying Kirchhoff's current law to the node of every subdomain. 
For a general system of $N$ subdomains this yields an implicit ordinary differential equation system of the form
\begin{subequations}
	\label{eq:thermalNetworkModel}
	\begin{equation}	
		\label{eq:thermalStateEq}
		\ODEMassMatrix\dot{\tempVec}(t) = \Amatrix\tempVec(t)+\powerSourceVec(t)
	\end{equation}
	with
	\begin{equation}
		\label{eq:systemMatrices}
		\begin{aligned}
			\ODEMassMatrix &\vcentcolon= 
			\begin{bmatrix}
				C_1 	& 			& 				& \\
						& C_2 	& 				& \\
						& 			& \ddots 	& \\
						& 			& 				& C_N
			\end{bmatrix}
			,\qquad \tempVec \vcentcolon=
			\begin{bmatrix}
				\temperature_1-\temperature_0\\
				\temperature_2-\temperature_0\\
				\vdots\\
				\temperature_N-\temperature_0
			\end{bmatrix}
			,\qquad \powerSourceVec = 
			\begin{bmatrix}
				P_1\\
				P_2\\
				\vdots\\
				P_N
			\end{bmatrix}
			,\\ 
			\Amatrix &= \diagMatrix+\Lambda ,\qquad [\Lambda]_{ij} = \Lambda_{ij}=\Lambda_{ji}\quad \text{for }i,j=1,\ldots, N,\\ 
			\diagMatrix &=
			\begin{bmatrix}
				-\Lambda_{10}-\sum_{j=1}^N\Lambda_{1j} 	& 																	& 				& \\
																				& -\Lambda_{20}-\sum_{j=1}^N\Lambda_{2j} 	& 				& \\
																				& 																	& \ddots 	& \\
																				& 																	& 				& -\Lambda_{N0}-\sum_{j=1}^N\Lambda_{Nj}
			\end{bmatrix}
			,
		\end{aligned}
	\end{equation}
\end{subequations}
cf.~\cite{Bac33}.
We emphasize that this may also be formulated as a differential-algebraic equation system if Kirchhoff's current law is explicitly incorporated as algebraic constraint.
Furthermore, these thermal equivalent circuit equations can be generalized by considering for instance temperature-dependent coefficients or source terms, cf.~\cite{BolK10,HolRF17,Ric67}.

\begin{remark}{\rm
    Usually, each of the $N$ subdomains is only connected by heat transfer to a few neighboring subdomains which results in a sparse matrix $\Lambda$ and, thus, also $\Amatrix$ is sparse. 
    In this case, the network can be characterized in a concise manner via the incidence matrix of the corresponding graph, see for instance \cite{SchM13}.}
\end{remark}

The parameters determining the thermal equivalent circuit equations \eqref{eq:thermalNetworkModel} can, for example, be determined based on the heat transfer theory, which relies on the geometry and material properties of the motor and possibly on empirical correlations for the convective heat transfer, cf.~\cite{BogCS05,QiSD14,SimWM14}.
Another approach is given by data-driven methods which estimate the parameters based on experimental data, see e.g.~\cite{BosP14,HubPB14,KraHL14,ZhuXLWT19}.
Also hybrid approaches combining analytical techniques and experimental data are common, see for instance \cite{NelGH21,SahRWPJ10,WalB15}.
Alternatively, some authors propose to use a numerical simulation based on the finite element method \cite{AlbB08} or based on the finite volume method \cite{JalTL08} for determining the coefficients of the thermal equivalent circuit.

For a thermal steady-state analysis, the capacitances and the time-dependency of the heat sources are neglected and the resulting steady-state equations take the form
\begin{equation*}	
	0 = \Amatrix\tempVec+\powerSourceVec,
\end{equation*}
where $\Amatrix$, $\tempVec$, and $\powerSourceVec$ are as defined in \eqref{eq:systemMatrices}, but without dependency on time.
Such a steady-state analysis based on thermal equivalent circuits has for instance been considered in \cite{BogCLP03,CheZC17,NerRP08,Rob69}.

\subsection{Model order reduction for the thermal simulation}

When applying a numerical discretization scheme as discussed in section~\ref{sec:thermal_discretization}, the resulting finite-dimensional model may  be of very high dimension, especially when a 3D model is considered.
Depending on the complexity of the thermal equivalent circuit, also the models arising from lumped-parameter modeling as addressed in section~\ref{sec:thermalNetworks} may be too expensive to be solved in real-time.
To overcome these difficulties, several model order reduction techniques have been applied in the literature.
In \cite{DemPAO10}, the authors first construct a thermal equivalent circuit model and formulate it as an input-output system, where the inputs correspond to the losses and the outputs are temperatures at certain nodes of interest.
Then, the system is discretized in time and the balanced truncation MOR scheme is applied to achieve a real-time simulation.
Also in \cite{Qi18} balanced truncation is used for reducing a thermal equivalent circuit model of an electrical machine.
A validation step demonstrates that  dimensions of the reduced order model of $10-20$  yield sufficiently accurate results.

In \cite{BerVTP00}, the authors compare two different MOR approaches. 
One is a variant of the method presented in \cite{EitBH87}, where the state of the reduced-order model corresponds to an approximation of the output of the full-order model (FOM).
The reduced system matrices are determined based on a minimization of the residual obtained when substituting the FOM output into the state equation of the ROM.
The other approach is based on a system identification technique proposed in \cite{PetHV97}, where a reduced model is determined in modal coordinates, i.e., the reduced system of ordinary differential equations is uncoupled.
The coefficient matrices are determined via an output error minimization based on simulated step response data.
The numerical results from \cite{BerVTP00} exhibit that the ROMs obtained by the identification method are faster and more accurate than the ones obtained by the method in \cite{EitBH87}.

In \cite{GaoCHH08}, the authors consider balanced truncation and another MOR technique which is based on pole-zero cancellations of the transfer function of the FOM.
They illustrate that both approaches yield low-dimensional and accurate surrogate models based on numerical experiments.
However, the considered FOM has only few state variables, although the authors argue that the framework may be also applied to systems of higher dimension.
In contrast, the authors of \cite{ZhoPH13} consider as FOM a three-dimensional finite element model, which is reduced via a modal truncation approach.
To this end, they split the set of eigenmodes into dynamic and static eigenmodes based on an excitation criterion.
Then, via neglecting the dynamics of the static eigenmodes, they only need to compute the dynamic eigenmodes and project the FOM onto their span.
The numerical experiments indicate that a ROM of dimension $30-40$ is sufficiently accurate and, moreover, allows for real-time simulations.

In contrast to the mechanical simulation addressed in section~\ref{sec:MORmechanics}, the simulation of the thermal full-order models is more often done based on self-implemented code, e.g., using \textsc{MATLAB}/\textsc{Simulink}, cf.~\cite{ChaZTF00,DemPAO10,Qi18,ZhoPH13}.
Nevertheless, there are also dedicated software packages for creating thermal models, see for instance \cite{BogCSSMM09} for an overview.

\section{Software implementation and data flow}
\label{sec:software}
The implemented  simulation models are integrated in the overall software architecture of the digital twin. The exploitation of the simulations within the digital twin requires it to communicate and to be connected with the master data from the digital world and with shadow data from the real world. Master data represent all the intrinsic properties of the electrical machine, while shadow data represent all the data that is measured by the sensors connected to the electrical machine. 

\subsection{Simulations in the overall software architecture of a digital twin}
\begin{figure}[H]
        \centering
        \includegraphics[scale=0.6]{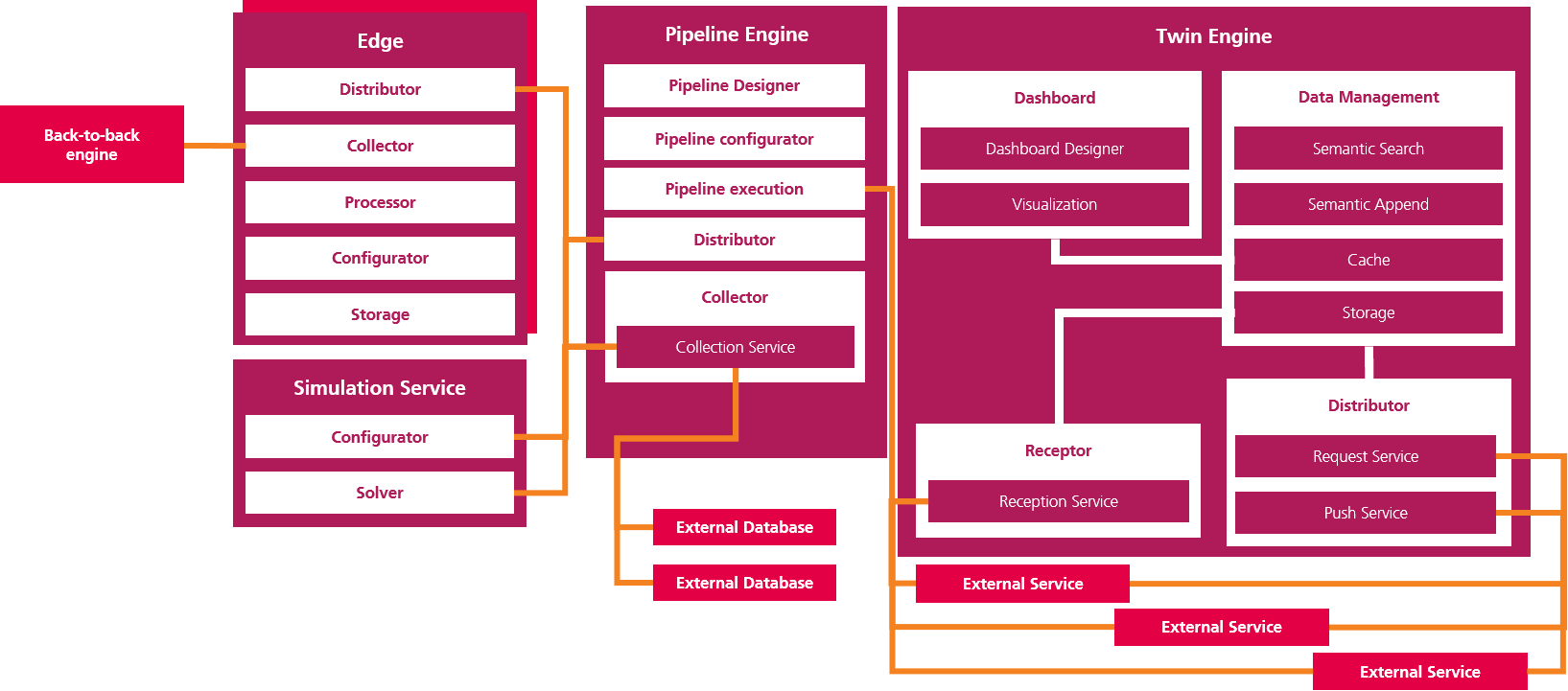}
        \caption{Software architecture for a digital twin of an electrical machine.}
        \label{fig:Architecture}
\end{figure}

The appropriate simulation model within the model hierarchy is chosen by the digital twin depending on the appropriate function and task. The software architecture required for these tasks is made up of various elements as shown in Figure \ref{fig:Architecture}.

At the center of the architecture is the Twin Engine, which takes over the functions of storing and displaying data. The external services are connected via the Receptor and Distributor. In addition, there are edge devices that record and pre-process the shadow data and make it available to the Twin Engine. Various services and databases are connected to the Twin Engine, including the simulation scripts. The connection between the shadow and master data outside and inside the Twin Engine is established by the pipelines in the Pipeline Engine. The pipelines in this project are programmed using \emph{NodeRed}. The integration of the simulation scripts via the pipelines within the framework of the digital twin will be discussed in greater detail in the next subsection.

In order to make the simulation scripts usable within the framework of the digital twin, they are connected to the architecture in the form of Dockers. In this project, the Dockers are hosted in an Amazon Web Services (AWS) cloud node and can communicate via defined interfaces. The communication of the input parameters for the scripts is realized using Message Queuing Telemetry Transport (MQTT) which is a network protocol for message queuing. To get the required inputs, the script subscribes to a specific topic and listens to all messages sent to the topic. If a message matches the required input parameters, the script is executed. The results are then published to another specific topic. The pipelines synthesize the data that the script needs to run the simulation. The pipelines aggregate the data from various sources into one message that is then used by the simulation. 
Once the simulation is performed, the results published by the simulations are read by the pipeline that assigns them to the corresponding assets in the asset structure in the Twin Engine. 
The various data can be visualized via the dashboarding in the Twin Engine. The pipelines are built using low-code building blocks in \emph{NodeRed}, so knowledge workers, such as engineers, can build these pipelines even without deep computer science knowledge. For more detailed information on the architecture we refer to the paper \cite{Lun23}.

\subsection{Data flow and pipelines}
\label{sec:Pipeline}
As an illustration, we consider the data flow required for the condition monitoring of electrical data. This is done by comparing real behavior coming from the digital shadow and theoretical behavior coming from the simulation results. 

\begin{figure}[H]
        \centering
        \includegraphics[scale=0.5]{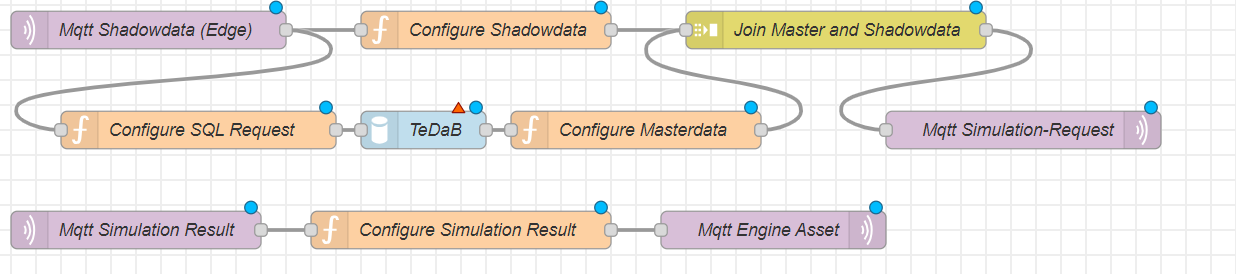}
        \caption{Simulation pipeline in NodRed.}
        \label{fig:Pipeline}
\end{figure}

The shadow data of the machine, in this case: current, voltage, electrical power, power factor, frequency, speed and torque, are recorded by the edge device. The edge device already offers the possibility to perform certain data pre-processing by means of edge computing. However, this is not considered further here. For more details check \cite{CAR19}. 
The data collected by the edge device is aggregated in a message and sent to the MQTT Broker. The pipelines listen to the corresponding topic over which the data of the edge device is transmitted as shown in Figure \ref{fig:Pipeline}. 

Since the simulation requires master data from the machine in addition to the shadow data, the pipelines must query various data sources and synthesize a corresponding message. The master data, such as specific resistances and reactances corresponding to the equivalent circuit parameters, are retrieved from the technical database using SQL queries as in Figure \ref{fig:Pipeline}. Subsequently, master and shadow data are sent in a common message to the simulation script in a specific topic. The simulation script receives this information and starts the calculation. The results, electrical data, efficiency and individual losses, are summarized in the form of another message containing the output parameters and metadata for unique identification with the input parameters. With the result topic, via which the script publishes the results, a pipeline transfers the data from this message to the corresponding topic and thus the corresponding asset, i.e., in our case, the corresponding electrical machine asset.

The simulation results are stored in the time series database of the TwinEngine. From there, they can be visualized in the dashboarding as discussed in section \ref{sec:Elements}. This enables condition monitoring of the shadow data and comparison with theoretical behavior, which is composed of shadow and master data.

\subsection{Software interface and visualization}
\label{sec:Elements}

\begin{figure}[H]
        \centering
        \includegraphics[scale=0.25]{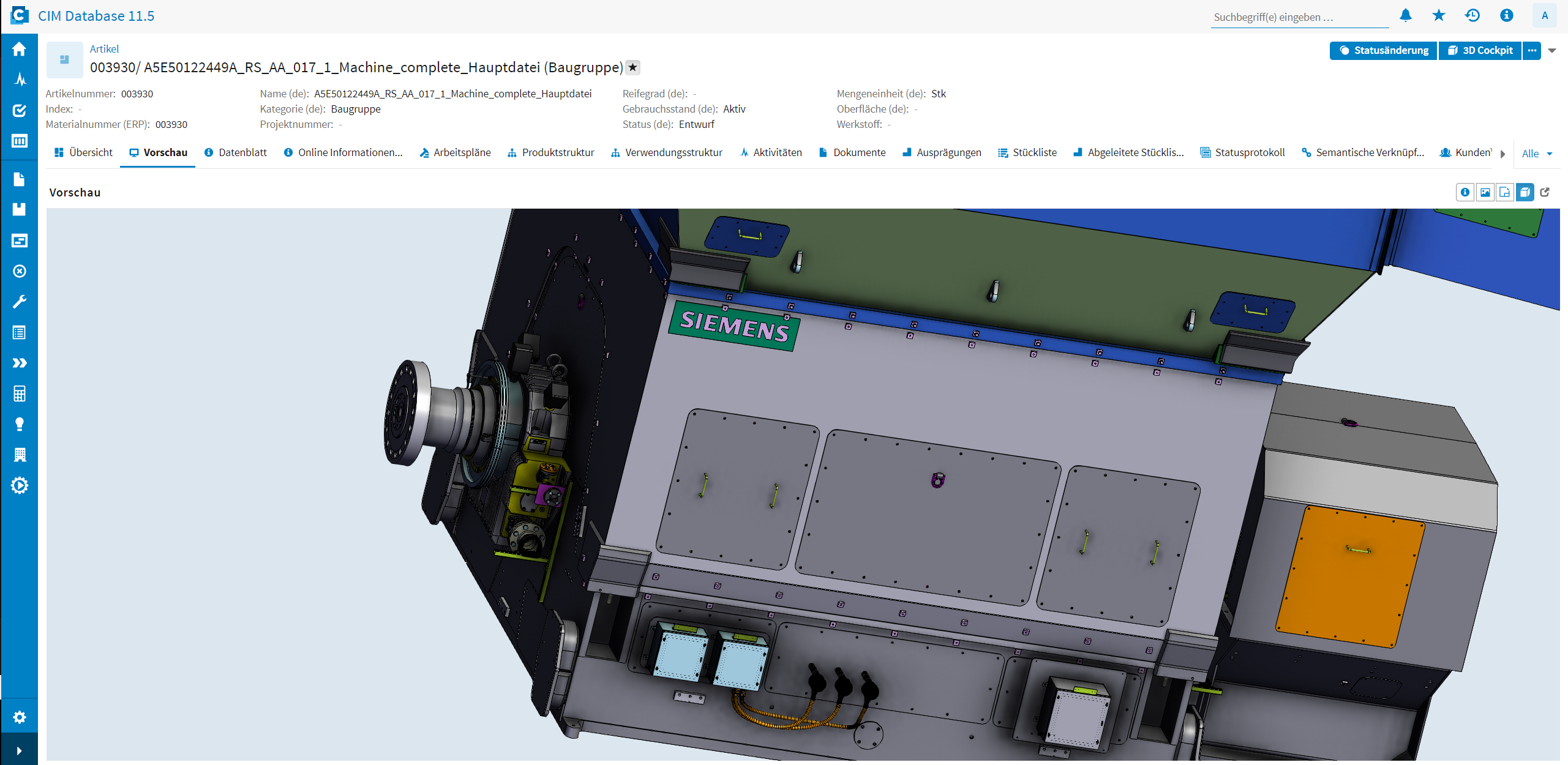}
        \caption{Web-based, interactive 3D visualization of the electrical drive in the PLM system as part of the digital twin operating system.}
        \label{fig:CEM}
\end{figure}

As depicted in Figure \ref{fig:Architecture}, the software architecture element "TwinEngine" is realized using the product life cycle management (PLM) software platform "\href{https://www.contact-software.com/en/products/iot-platform-for-digital-business-models}{CONTACT Elements for IoT}" (CE4IoT) which is designed to create and operate digital twins in complex industrial scenarios. CE4IoT is interconnected to models, data and information that has been created along the product life cycle - from engineering to production, to operation and vice versa. The 3D model of the electrical machine can be uploaded to CE4IoT along with the relevant data from the digital shadow and simulation as shown in Figure \ref{fig:CEM}.

\begin{figure}[H]
        \centering
        \includegraphics[scale=0.32]{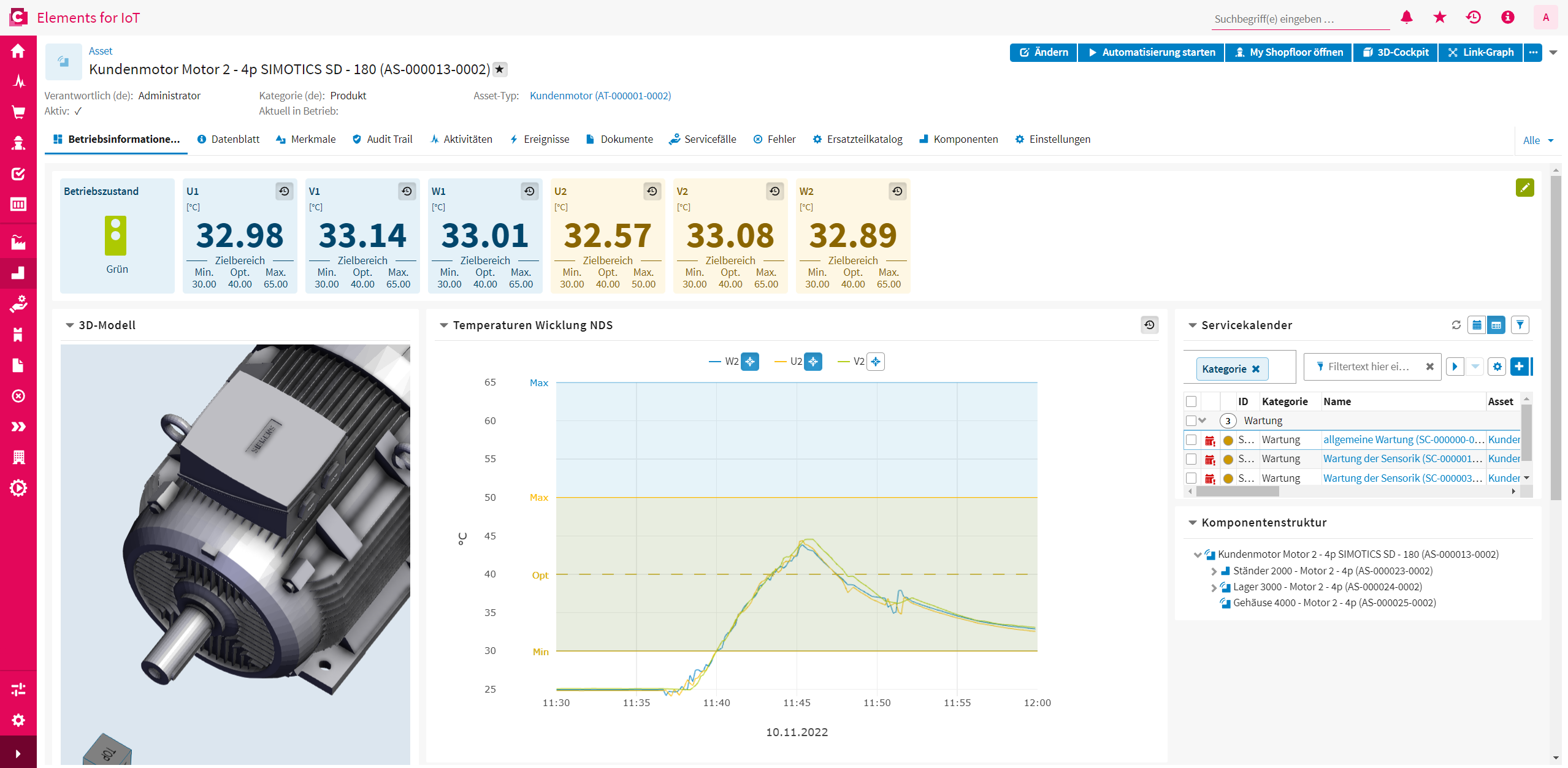}
        \caption{Dashboard of a digital twin of an electrical machine in CONTACT Elements for IoT.}
        \label{fig:Dashboard}
\end{figure}

During operation, the data coming from the sensors and simulation data are stored in the time series database. These can be visualized in the dashboard of CE4IoT as shown in Figure \ref{fig:Dashboard}. One can see the instantaneous values as well as a history of the field data (digital shadow), simulation data, and threshold values in a time series graph.
The health status is displayed as a traffic light display for the condition monitoring part. Green means that the machine is in a healthy state. Yellow means that the machine may be in danger (in yellow above 10\% deviation) and finally red means that there is 
an imminent danger  (in red above 20\% deviation). In addition, in case of deviation, an automatic message is sent to the person responsible. The stored data can be also provided for download and evaluation in third-party software.

\section{Implementation: demonstrator}
\label{sec:demonstrator}
\begin{figure}[H]
        \centering
        \includegraphics[scale=0.6]{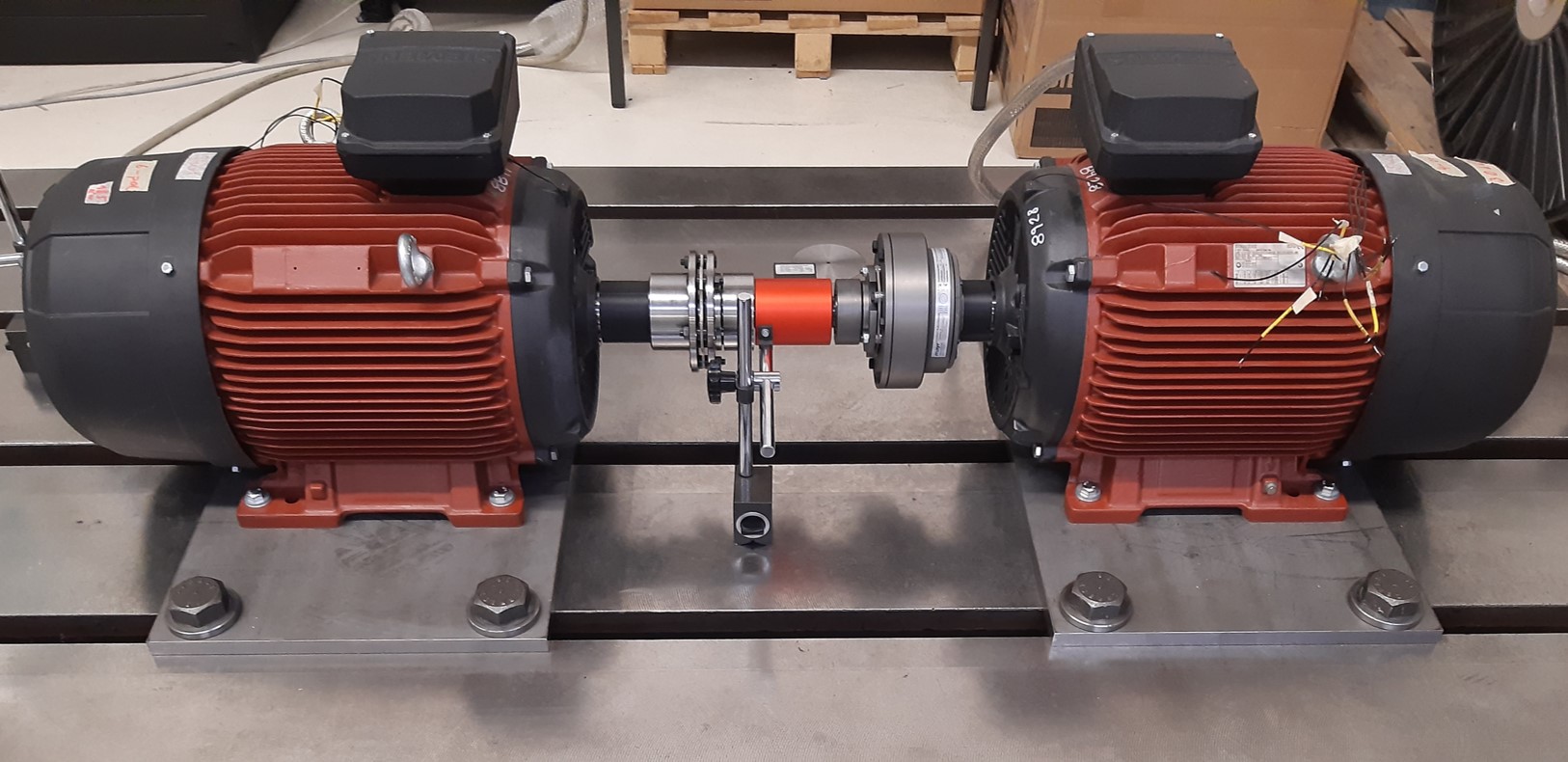}
        \caption{Physical demonstrator: Two electrical machines coupled by a drivetrain.}
        \label{fig:Demo1}
\end{figure}

The concept for the digital twin and the data flows described in section \ref{sec:software} were implemented using a hardware demonstrator in combination with an implemented software demonstrator to test the developed structure and the various services. The hardware demonstrator consists of two asynchronous electric motors facing each other and coupled by a drive-train as shown in Figure \ref{fig:Demo1}.

\subsection{Technical details about the demonstrator}
The demonstrator consists of two motors where one symbolizes the customer motor and the other a test field machine. The drive-train consists of a torque measuring shaft and a slip clutch that protects the motors from overloads. In the first design, a flexible coupling was to be used, whose service life was to be predicted in combination with a load model. For safety reasons, this design was abandoned and a safety clutch was used, which can be used to predict the service life of the friction linings using a friction lining wear model.

\begin{figure}[H]
        \centering
        \includegraphics[scale=0.6]{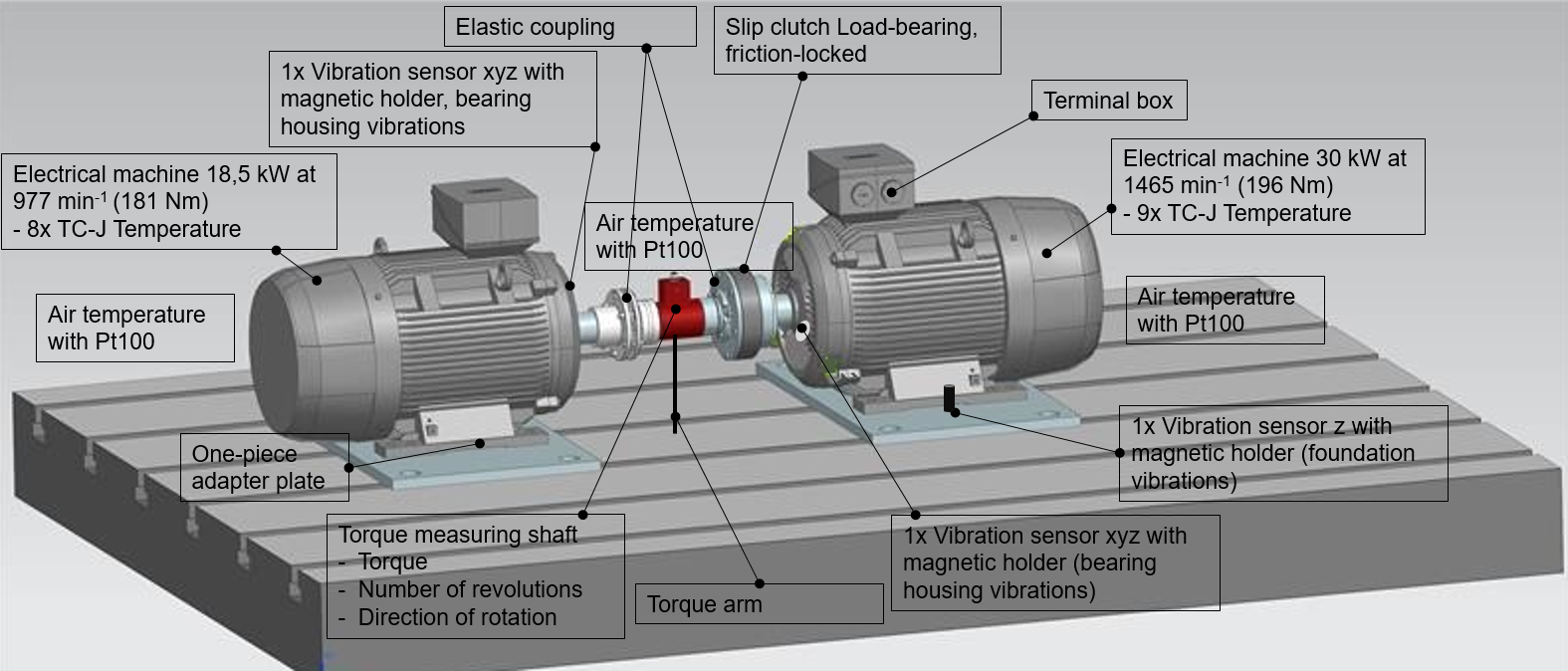}
        \caption{Physical demonstrator: sensor placement.}
        \label{fig:Demo_sensor}
\end{figure}

 The machines are supplied with sensors that enable the implementation of the digital twin as shown in Figure \ref{fig:Demo_sensor}.
 The customer's motor is a 4-pole machine with 30 KW which, at the operating point, has 196 Nm of torque at a speed of 1465 $min^{-1}$(rpm). At the same operating point, the power factor is 0.81. This machine is supplied with 11 thermocouples, which record the temperatures of the machine at bearings, windings and housing.

The test field motor is a 6-pole machine with 18.5 KW which has 181 Nm torque at a speed of 977 $min^{-1}$(rpm) at the operating point with a power factor of 0.77. For thermal measurements, 8 thermocouples record the temperatures of the machine at bearings, windings and housing.
In addition to the thermocouples installed in the machines, there are three PT-100 sensors to measure the ambient temperatures, so that the cooling air inlet and outlet temperatures are recorded. Also, via the torque measuring shaft, the speed and torque are recorded and passed on to the inverter. This also supplies the other electrical operating data of the machines.

The signals from the thermocouples are passed on to the TresCOM-Edge and RapsberryPI edge devices. The edge devices evaluate the signals and send them via MQTT to the software demonstrator using the data flow described in section \ref{sec:software}.
As described in section \ref{sec:software}, the software demonstrator consists of the CE4IoT platform, dockers deployed in Fargate in an Amazon Web Services(AWS) server, and other services and databases connected via defined interfaces.

\begin{figure}[H]
        \centering
        \includegraphics[scale=0.38]{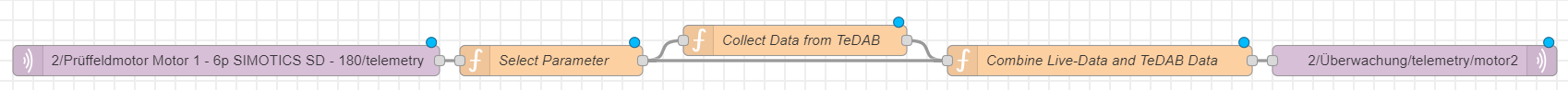}
        \caption{Pipeline of simulation data for demonstrator}
        \label{fig:Pipeline_demo}
\end{figure}

The MQTT broker manages the MQTT communication through which the edge devices, pipeline engine, asset management, and services communicate with each other. The database for the simulated systems is connected to the pipelines using a WebDAV interface. Other technical databases are connected via SQL interface.
Therefore, the pipelines can retrieve and synthesize data from different systems via the defined interfaces. The external services are then addressed via the pipelines, thus fulfilling the added values behind the user stories.
The data coming from the shadow data and the master data stored in databases are used to set up the implemented simulation in a pipeline as shown in Figure \ref{fig:Pipeline_demo}. 
In the case of condition monitoring, the results of the simulation are used to check that there are no faults in the operation. As an illustration, we show how the digital twin is used in the context of condition monitoring by comparing measured data from an experiment conducted on the demonstrator and results from the simulation engine within the digital twin.

\subsection{Experimental results}
\label{sec:experiment}

The conducted experiment consists of running the demonstrator in different operating points including constant frequency, changing frequency, and changing flux. A scenario where a fault is provoked is also considered. 
During operation the data needed to setup the simulation is sent to the simulation engine, the simulation is performed and the simulation results are sent back. These are compared with shadow data from the demonstrator and displayed in CE4IoT. If there is a significant deviation an error message is displayed. The simulation is updated each second with the new operating conditions. Two different experiments were conducted with different operating conditions in each experiment.
The first experiment was conducted with the 4-pole machine as the motor. A summary of the comparison results of shadow data and simulation data in some operational conditions is presented in Table \ref{Table_4pole}. Note that in these scenarios we are not operating in nominal frequency and voltage. As a result, some calculations from section \ref{sec:equ_ckt} have to be adapted with a frequency ratio that is equal to the operating frequency divided by the nominal frequency. The nominal frequency in this case is $50Hz$. 
Particularly, the reactances in the equivalent circuit of section \ref{sec:equ_ckt} have to be scaled with the frequency ratio. 
A second experiment was conducted with the 6-pole machine as the motor. The results are summarized in Table \ref{Table_6pole}.

\begin{table} [!h]
\begin{center}
\begin{tabular}{ ||c|| c | c |c|| c | c | c || c | c | c ||}

 Scenarios & \multicolumn{3}{c||}{\shortstack{Scenario 1} } & \multicolumn{3}{c||}{\shortstack{Scenario 2 } } & \multicolumn{3}{c||}{\shortstack{Scenario 3 } } \\ 
 
 Data type & Measured & sim & err & Measured & sim & err & Measured & sim & err  \\ 
 
 Voltage (V) & 256 & 256 & 0\% & 288 & 288 & 0\% & 317 & 317 & 0\%  \\ 
 
 Frequency (Hz) & 35.8 & 35.8 & 0\% & 35.5 & 35.5 & 0\%  & 35.5 & 35.5 & 0\%  \\  
 
 Power load (kW) & 20.6 & 20.6 & 0\% &  20.7 & 20.7 & 0\% & 20.9 & 20.9 & 0\% \\  
 
 Current (A) & 60.2 & 58.4 & 3\% & 56.7 & 54.6 & 3.7\% & 56.8 & 56.5 & 0.5\% \\   
 
 Power factor & 0.86 & 0.89 & 3\% & 0.82 & 0.84 & 2\% & 0.74 & 0.74 & 0\% \\     
 Torque (N.m) & 191 & 191 & 0\% & 191 & 192 & 0.5\% & 191 & 192 & 0.5\% \\ 
 
 Speed (rpm) & 1035 & 1028 & 0.6\% & 1035 & 1030 & 0.5\% & 1035 & 1036 & 0.1\%   

\end{tabular}
\caption{\label{Table_4pole}Measured data, simulation (sim) data and the relative error between them for 4-pole motor.}
\end{center}
\end{table}

\begin{table} [!h]
\begin{center}
\begin{tabular}{ ||c|| c | c | c|| c | c | c ||}

 Scenarios & \multicolumn{3}{c||}{\shortstack{Scenario 1 } } & \multicolumn{3}{c||}{\shortstack{Scenario 2} }  \\ 
 
 Data type & Measured & sim & err & Measured & sim & err  \\ 
 
 Voltage (V) & 398 & 398 & 0\% & 372 & 372 & 0\% \\ 
 
 Frequency (Hz) & 50.4 & 50.4 & 0\% & 50.6 & 50.6 & 0\%  \\  
 
 Power load (kW) & 19.2 & 19.2 & 0\% & 19.1 & 19.1 & 0\% \\  
 
 Current (A) & 40 & 38 & 5\% & 41.2 & 38.9 & 5\%\\   
 
 Power factor & 0.8 & 0.79 & 1\% & 0.82 & 0.82 & 0\%\\     
 Torque (N.m) & 188 & 186 & 1\% & 186 & 185 & 0.5\%\\  
 
 Speed (rpm) & 980 & 983 & 0.3\% & 980 & 984 & 0.4\%   

\end{tabular}
\caption{\label{Table_6pole}Measured data, simulation (sim) data and the relative error between them for the 6-pole motor.}
\end{center}
\end{table}

In each scenario, the measured voltage, frequency and power load are transmitted to the simulation engine to compute: the current, power factor, torque speed and losses. In this case, the simulation uses the lumped parameter model described in section \ref{sec:equ_ckt} as it is the fastest and thus most suitable for condition monitoring. The results of the simulation (sim) are compared to the measured values from the digital shadow and the relative error is computed 
\[err=\frac{\lvert measured-sim \rvert}{measured}.\]

The highest relative error is in the order of $5\%$ which is good enough for the condition monitoring task as the lowest fault threshold is $10\%$. For other functions of the digital twin, a more accurate model has to be used in the model hierarchy. 

\begin{figure}[H]
        \centering
        \includegraphics[scale=0.6]{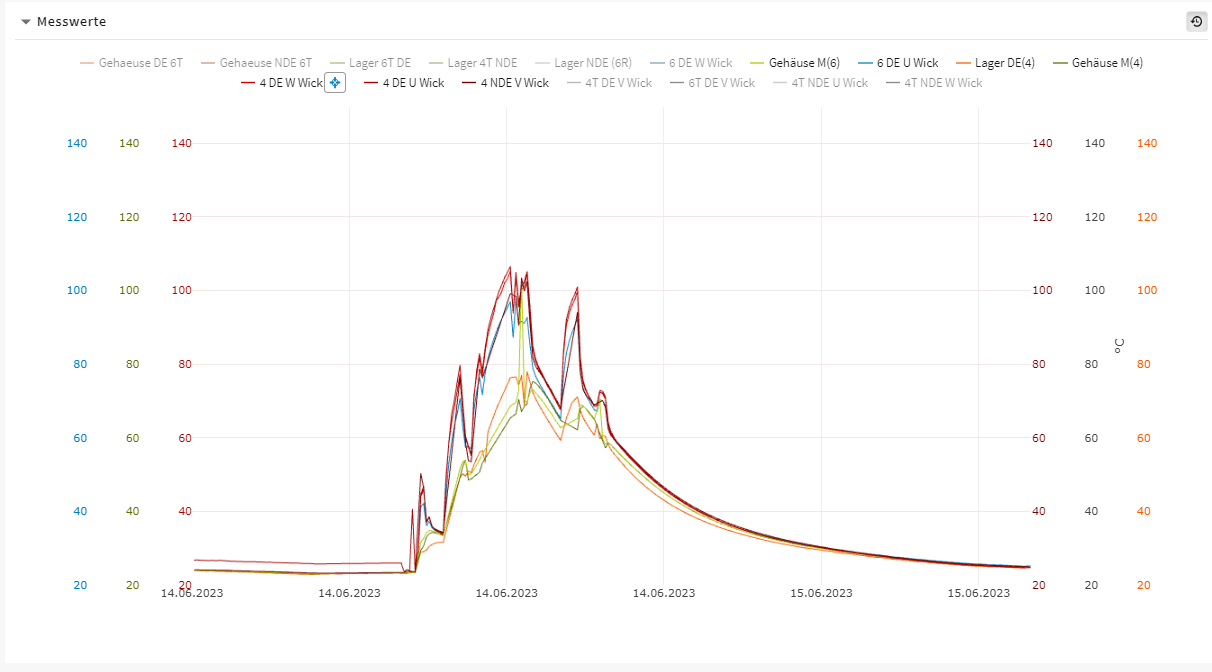}
        \caption{Temperature measurements in function of time as seen in CE4IoT.}
        \label{fig:Temp_demo}
\end{figure}

If there is any significant deviation from normal operation, the digital twin will detect the fault and display a message in the software interface. In addition to the data that can be measured, the simulation is also able to provide information about losses, efficiency, stator current and rotor current as discussed in section \ref{sec:electricalSim}. This data is stored in a database and can always be consulted from the dashboard of the software interface. The stored data can then be used for the optimization of machine operation and the improvement of future designs. Temperature data is also stored in the database. 
The temperature measurements recorded over time during the experiment can be seen in Figure \ref{fig:Temp_demo} as a time series graph displayed in the dashboard of the CE4IoT platform. 
Each color in the graph represents one temperature sensor. One can clearly see in Figure \ref{fig:Temp_demo} the startup point and then the switching between different operation conditions. In the end, we see the cooling down of the electrical machine after it is turned off. 
The temperature measurements shown in Figure \ref{fig:Temp_demo} are compared with a threshold set by the simulation based on the particular operating condition. 
The threshold differs depending on the placement of the sensor and the expected temperature in the place where the sensor is placed.

The experimental results show that a weak coupling between different domains provides results that are satisfactory for the condition monitoring function of the digital twin. However for more accurate future designs, one has to consider a full coupled model between the electromagnetic, mechanical and thermal domains. This is out of the scope of this paper and will be considered in future work.

\section{Conclusion}
In this paper, we have presented the main simulation tasks in a digital twin of an electrical machine. 
These kinds of simulations have been already discussed in the literature for decades.
However, now the quest for digital twins brings new requirements and challenges. 
We have also discussed some of the approaches to overcome these challenges.
In particular, it is pointed out that model reduction is a crucial tool for real-time simulations and model hierarchies. 
In close relation to that is surrogate modeling that uses sensor data gathered by the digital twin to improve lifetime models used by the digital twin.  As an illustration, we showed how a digital twin can be implemented in an industrial context with a physical demonstrator. The simulation and software architecture were implemented for the demonstrator and the experimental results show successfully how the digital twin is used in practice.

This paper shows how a digital twin can be implemented in the industrial context of electrical machines. However, there are still points that need to be improved in order to achieve a full digital twin as discussed in this paper. These points can be summarized as follows:
\begin{itemize}
    \item Build a catalogue of models including a hierarchy from the most fundamental equations to the simplest and fastest models. Model reduction is a key tool in this context.
    \item Achieve a full multiphysics coupling between electromagnetic, mechanical and thermal domains.
    \item Develop a complete software solution that allows for an efficient data flow, data visualization and adaptive and hierarchical modeling within the digital twin. This may include a combination of existing software solutions.
    \item Create an automated machinery that can produce machine-specific models based solely on machine data.
    \item Build life cycle models and adaptive models over time using data assimilation and data-driven modeling in order to improve condition monitoring and predictive maintenance.
\end{itemize}

\subsection*{Acknowledgment}
The authors acknowledge funding from ProFIT (co-financed by 'Europ\"{a}ischen Fonds f\"{u}r regionale Entwicklung' (EFRE)) within the WvSC project: EA 2.0 - Elektrische Antriebstechnik.
The authors would like to thank Karsten Brach from Siemens AG for his valuable comments and discussions. The authors would like also to thank Hartmut Rauch from Siemens AG and Thomas Damerau and Eric Storz from CONTACT Software GmbH for the discussions and contributions to the EA 2.0 project.

\bibliographystyle{alpha}
\bibliography{main}

\end{document}